%% file: HyperLTL_arXiv.tex
\documentclass[reqno]{amsart}
\input{header}

\begin{document}

\title[Formal Verification of Uncertain Systems against Hyperproperties]{Verification of Hyperproperties for Uncertain Dynamical Systems via Barrier Certificates}

\author{Mahathi Anand$^{1}$}
\author{Vishnu Murali$^{2}$}
\author{Ashutosh Trivedi$^{2}$}
\author{Majid Zamani$^{2,1}$}
\address{$^1$Department of Computer Science, LMU Munich, Germany.}
\email{mahathi.anand@lmu.de}
\address{$^2$Department of Computer Science, University of Colorado Boulder, USA.}
\email{vishnu.murali@colorado.edu,ashutosh.trivedi@colorado.edu,majid.zamani@colorado.edu}

 \maketitle       

\begin{abstract}
\input{abstract}
\end{abstract}

\section{Introduction}
\label{sec:introduction}
\input{introduction}

\section{Problem Definition}
\label{sec:prelims}
\input{prelims}

\section{Augmented Barrier Certificates (ABCs)}
\label{sec:labeling}
\input{problem}

\section{ABCs via Sum-of-Squares Programming}
\label{sec:sos}
\input{synthesis}

\newpage

\section{Case Studies}
\label{sec:case-studies}
\input{casestudies_new}

\section{Discussion and Conclusion}
\label{sec:discussion}
\input{discussion}

 \bibliographystyle{alpha}
 \bibliography{HyperLTL_arXiv.bbl}

\section{Appendix}
\label{dec:appendix}
\input{appendix}
\raggedbottom

\end{document}

%% file: header.tex
\usepackage{multicol}
\usepackage{amsmath}
\usepackage{temporal}
\usepackage{graphicx}
\usepackage[misc,geometry]{ifsym}
\usepackage{hyperref}
\usepackage{amssymb,latexsym,amsfonts,amsmath}
\usepackage{caption,subcaption}
\usepackage{graphicx}
\usepackage{paralist}
\usepackage{dsfont}
\usepackage{booktabs}
\usepackage{eso-pic}
\usepackage{epsfig}
\usepackage{url}
\usepackage{epstopdf}
\usepackage{tikz}
\usetikzlibrary{automata,positioning}
\usepackage{bm}
\usepackage{verbatim}

\usepackage{color}

\usepackage{etoolbox}
\AtBeginEnvironment{thebibliography}{\linespread{1}\selectfont}

\DeclareSymbolFont{symbolsC}{U}{pxsyc}{m}{n}
\SetSymbolFont{symbolsC}{bold}{U}{pxsyc}{bx}{n}
\DeclareFontSubstitution{U}{pxsyc}{m}{n}
\DeclareMathSymbol{\medcirc}{\mathbin}{symbolsC}{7}
\usetikzlibrary{arrows,calc,decorations.markings,math,arrows.meta}
\usepackage{algorithm}
\usepackage{algorithmicx}
\usepackage[noend]{algpseudocode}
\makeatletter
\newcommand{\ALOOP}[1]{\ALC@it\algorithmicloop\ #1%
  \begin{ALC@loop}}
\newcommand{\ENDALOOP}{\end{ALC@loop}\ALC@it\algorithmicendloop}

\makeatother\algnewcommand{\Initialize}[1]{%
	\State \textbf{Initialize:}
	\Statex \hspace*{\algorithmicindent}\parbox[t]{.8\linewidth}{\raggedright #1}
}
\numberwithin{equation}{section}

\def\set#1{{\{#1 \}}}

\newcommand{\R}{{\mathbb{R}}}

\newcommand{\B}{{\mathcal B}}

\newcommand{\N}{{\mathbb{N}}}

\newcommand{\bfm}{\mathbf}

\newcommand{\Ss}{\mathfrak{S}}

\newcommand{\AP}{\mathsf{AP}}

\DeclareMathOperator{\infi}{inf}
\newcommand{\zip}{\textsf{zip}}
\newcommand{\unzip}{\textsf{unzip}}
\newcommand{\dtCS}{system }

\newtheorem{theorem}{Theorem}[section]
\newtheorem{lemma}[theorem]{Lemma}

\newtheorem{problem}[theorem]{Problem}

\newtheorem{definition}[theorem]{Definition}
\newtheorem{example}[theorem]{Example}

\newtheorem{remark}[theorem]{Remark}
\newtheorem{assumption}[theorem]{Assumption}
\numberwithin{equation}{section}

%% file: abstract.tex
Hyperproperties are system properties that require quantification over multiple
execution traces of a system. 
Hyperproperties can express several specifications of interest for
cyber-physical systems---such as opacity, robustness, and
noninterference---which cannot be expressed using linear time properties. 
This paper presents for the first time a discretization-free approach for the formal
verification of discrete-time uncertain dynamical systems against hyperproperties. 
The proposed approach involves decomposition of complex hyperproperties into
several verification conditions by exploiting the automata-based structures
corresponding to the complements of the original specifications. 
These verification conditions are then discharged by synthesizing so-called \emph{
augmented barrier certificates}, which provide certain safety guarantees for the
underlying system.  
For systems with polynomial-type dynamics, we present a sound procedure to synthesize
polynomial-type augmented barrier certificates by reducing the problem to
sum-of-squares optimizations. 
We demonstrate the effectiveness of our proposed approaches on two physical
case studies against two important hyperproperties: \emph{initial-state opacity} and
initial-state robustness. 

%% file: introduction.tex
Classical control theory provides theory, techniques, and tools to analyze {\it complex} dynamical
systems against {\it simple} objectives such as stability. 
Traditional formal methods, on the other hand, tend to focus on developing approaches to verify software/hardware systems with  \emph{simpler} discrete dynamics against structurally \emph{rich} logic-based specifications concerning safety and liveness.  
Cyber-physical systems (CPS)---systems characterized by structured software systems interacting with equally complex physical systems---blur this traditional demarcation between control systems and software systems.
The critical role of CPS in modern society in safety- and security-critical applications has spurred interest in developing approaches to provide rigorous guarantees for such systems.
As a result, in the last two decades, formal verification of CPS with complex continuous state-space dynamics against logic-based specifications has received considerable attention~\cite{tabuada2009verification,belta_formal_2017}.

Verification of CPS has been typically performed against temporal logic specifications, notably linear temporal logic (LTL) and computation tree logic (CTL) \cite{baier2008principles}, expressing properties of a set of desirable system executions.  
While these logics can describe a large number of specifications of interest
that consider individual execution traces of CPS, many important information-flow
properties and planning objectives involve relating multiple execution traces. 
These properties cannot be expressed by classical temporal logic specifications equipped to express properties of individual traces.
To fill this gap, Clarkson and Schneider~\cite{clarkson_hyperproperties_2010} introduced the notion of \textit{hyperproperties} as properties of collective behavior relating multiple execution traces. 

As an example of a hyperproperty, consider a security property in a CPS prone to \emph{intrusion attacks}.
Suppose that this system requires that secret information is never revealed,
\textit{i.e.}, observations from the outside remain indistinguishable from each
other, despite the secret. This specification, known as
opacity \cite{Mazare04usingunification}, requires us to relate and quantify two
observation traces simultaneously. Similarly, an optimality objective
\cite{wang_hyperltl} for a robotic system would require the existence of a trace
that is more favorable than all the other traces of the system, again
quantifying multiple execution traces at a time. Other examples for
hyperproperties include noninterference \cite{noninterference} and observational
determinism \cite{obdet}. In order to formally specify hyperproperties,
the hyper-temporal logic HyperLTL was introduced in \cite{clarkson_temporal_2014}. 
HyperLTL, developed as an extension to LTL, uses
trace variables to denote individual execution traces and utilizes universal
($\forall$) and existential ($\exists$) quantifiers before a quantifier-free formula
over atomic propositions to specify on which traces the atomic propositions must
hold. 

Formal verification of hyperproperties has been studied in the context of
finite-state systems. However, the techniques
used for finite systems are not applicable to real-world CPS which evolve on 
continuous (or even hybrid) state spaces. 

\vspace{0.5em} \noindent \textbf{Contributions.}  We aim at bridging this gap by
presenting a discretization-free, systematic, and sound verification procedure
based on a notion of barrier certificates for discrete-time uncertain dynamical systems
against hyperproperties. In particular, we consider those specifications that
can be expressed by HyperLTL formulae~\cite{clarkson_temporal_2014}.  The verification procedure is achieved by decomposing the given specification into simpler safety tasks, so-called \textit{conditional invariance},  by constructing an implicitly quantified  B\"uchi automaton corresponding to the complement of the specification. We
introduce {\it augmented barrier certificates} (ABCs), defined over an augmented
system obtained by taking the product of the original system
with itself (self-composition), which provide us with sufficient conditions ensuring the
satisfaction of those conditional invariances. 

 Inspired by the results in~\cite{wongpiromsarn_automata_2016}, we propose an automata-theoretic approach to extend the applicability of ABCs beyond conditional invariance to HyperLTL specifications by finding barrier certificates ensuring the possibility of avoiding accepting traces of corresponding automata by disallowing certain transitions on different lassos (simple path followed by a simple accepting cycle). To do so, ``existential'' player is required  
to select a trace before knowing the choices of the ``universal'' player. This necessitates the need for a 
common ABC for some transitions of all lassos, which may be hard to ensure in practice.
On the other hand, when the HyperLTL property belongs to $\forall^{*} \exists^{*}$-fragment \cite{clarkson_temporal_2014}, we can exploit a similar approach as in~\cite{wongpiromsarn_automata_2016} to use separate ABCs to provide the necessary guarantees by leveraging the structure of the automata corresponding to the negation of specifications. 
For systems with polynomial-type dynamics, we present a
sum-of-squares (SOS) approach to compute polynomial-type ABCs for the individual
conditional invariance. Finally, we demonstrate the effectiveness of our proposed
approach by verifying two physical case studies with respect to initial-state
opacity and initial-state robustness, respectively, which both can be described
by HyperLTL formulae. Proofs of all statements are provided in the Appendix.

\vspace{0.5em} \noindent \textbf{Related Literature.} There have been several results in the literature
for the verification and synthesis of CPS against temporal logic specifications.
Many earlier results have utilized abstraction-based techniques based on
state-space discretization. Examples include abstraction-based framework for
linear systems \cite{tabuada_ltl}, for nonlinear systems \cite{majid}, synthesizing feedback strategies for
piece-wise affine systems \cite{Yordanov}, and counterexample-guided abstraction
refinement (CEGAR) for nonlinear systems \cite{wolff_cegar} to name a few. More
recently, automata-theoretic, discretization-free approaches via barrier
certificates \cite{prajna_safety_2004} have been utilized for the verification
of LTL specifications in the context of nonlinear systems
\cite{wongpiromsarn_automata_2016}, hybrid systems \cite{bisoffi}, as well as
stochastic systems \cite{jagtap_temporal_2018,anand_verification_2019}.  

Unfortunately, most of the existing results pertaining to hyperproperties are
tailored to finite-state transition systems. For example,
the results in  \cite{Finkbeiner2015AlgorithmsFM} present a practical verification approach for
finite-state systems with respect to alternation-free fragments of HyperLTL formulae. The proposed approaches in \cite{clarkson_temporal_2014} present a model-checker for HyperLTL
specifications with alternation depth of at most one. The results in \cite{finkbeiner_model_2018} propose a new model checking algorithm based on
model-counting for quantitative hyperproperties. A bounded model
checking algorithm for hyperproperties is proposed in \cite{hsu2020bounded}.
Verification of other types of hyperproperties such as $k$-safety
hyperproperties and hyperliveness properties have also been studied in \cite{Canonicalk-safety} and \cite{VerifyingHyperLiveness}, respectively. Checking satisfiability of
certain fragments of HyperLTL specifications, such as the ``$\forall^{*}
\exists^{*}$" fragment, are undecidable in general
\cite{finkbeiner_et_al:LIPIcs:2016:6170}. 
Formal verification of continuous state-space
CPS against general hyperproperties remains largely unexplored. 
Hyperproperties have been studied for CPS in \cite{viethyperSTL} as well as \cite{statistical_verification_hyper}, but in the context
of falsification and statistical model checking, respectively. These results are empirical and rely on experimental simulations, and therefore do not provide sound guarantees. Finally, we would like to mention that an extended abstract of this work was presented in~\cite{Extended_Abstract_CAADCPS2021}.

%% file: prelims.tex
We write $\R$ and $\N$ to denote the set of real and natural numbers, respectively.
Appropriate subscripts are used to restrict the sets, e.g., $\R_{>0} = \set{x
  \in \R  > 0}$ denotes the set of positive reals. 
We write $\mathbb{R}^n$ to denote $n$-dimensional Euclidean space equipped with
Euclidean norm $\|x\|$.
For a finite set $A$, the cardinality of $A$ is denoted by $|A|$. 

For a family $x_1 \in \mathbb{R}^{n_1}, x_2 \in
\mathbb{R}^{n_2}, \ldots, x_N \in \mathbb{R}^{n_N}$ of $N$ vectors, we write $(x_1, x_2,
  \ldots, x_N)$ to denote the corresponding vector of dimension $\sum_i n_i$.
For a set $A$, we write $A^n$ for the $n$-ary Cartesian power of $A$,
i.e. $A^n=\set{(a_1,a_2,\ldots,a_n) \mid a_i \in A \text{ for all } 1 \leq i
  \leq n}$.  
For a tuple $t = (a_1, a_2, \ldots, a_k)$ and $1 \leq i \leq k$ we write $t(i)$ for its $i$-th element 
and $t_{\leq i}$ for the tuple $(a_1, a_2, \ldots, a_i)$. 
Similarly, for a mapping $f : A \to B$, we define its $n$-ary Cartesian power $f^n: A^n
\rightarrow B^n$, as  $(a_1, \ldots, a_n) \mapsto (f(a_1), \ldots, f(a_n))$.

An alphabet $\Sigma$ is a finite set of letters. 
An $\omega$-sequence $\bm{\sigma} = \textsc{w}_0\textsc{w}_1\ldots$ is an infinite concatenation 
of letters, i.e. for all $i \geq 0$ we have $\textsc{w}_i \in \Sigma$.
A finite sequence is such a sequence but with a finite length.
We write $\Sigma^*$ and $\Sigma^\omega$ for the set of finite and $\omega$-sequences over $\Sigma$, 
and we let $\Sigma^\infty = \Sigma^* \cup \Sigma^\omega$.
For a sequence $\bm{\sigma} = \textsc{w}_0\textsc{w}_1\ldots \in \Sigma^\omega$, let
$\bm{\sigma}[i]$ be the $i$-th element $\textsc{w}_i$ of $\bm{\sigma}$ and $\bm{\sigma}[i,\infty]$ for the
$\omega$-sequence $\textsc{w}_i\textsc{w}_{i+1}\ldots \in \Sigma^\omega$ of $\bm{\sigma}$ starting from $i$-th position.

We let $\zip: (\Sigma^\omega)^p \to (\Sigma^p)^\omega$ denote a function that maps a $p$-tuple of sequences to a
sequence of $p$-tuples, i.e.
\begin{equation*}
  (\bm{\sigma_1}, \bm{\sigma_2}, \ldots, \bm{\sigma_p}) \to (\bm{\sigma_1}[0], \bm{\sigma_2}[0], \ldots, \bm{\sigma_p}[0])(\bm{\sigma_1}[1], \bm{\sigma_2}[1], \ldots, \bm{\sigma_p}[1]), \ldots, 
\end{equation*}
and $\unzip: (\Sigma^p)^\omega \to (\Sigma^\omega)^p$ denotes the inverse of $\zip$, i.e. 
\begin{align*}
  \bm{\sigma}  \to &( 
    (\bm{\sigma}[0](1) \bm{\sigma}[1](1) \ldots),
    \ldots, (\bm{\sigma}[0](p) \bm{\sigma}[1](p) \ldots)).
\end{align*}

\subsection{Discrete-Time Uncertain Dynamical Systems}

  A discrete-time uncertain dynamical system (or simply, system) is a tuple $\Ss = (X,W,f)$,
   where $X \subseteq \mathbb{R}^n$ and $W \subseteq
  \mathbb{R}^m$ are the (potentially uncountable) state and exogenous input sets, and $f: X \times W
  \rightarrow X$ is the transition function that characterizes the state
  evolution. The evolution or \emph{run} of the system $\Ss$ for a given initial state $x_0 \in
 X$  and exogenous input sequence $\nu: \N \rightarrow W$, denoted by $\bfm{x}_ {(x_0,\nu)}$, is given by 
 sequence  $\bfm{x}:\N {\rightarrow} X$, where 
 \begin{eqnarray}
   \bfm{x}(t) = \begin{cases}
     x_0 & \text{ if $t = 0,$} \\ 
     f(\bfm{x}(t-1),\nu(t-1)) & \text{ otherwise.}  
   \end{cases}
   \label{eq:dyn}
\end{eqnarray}

For a system $\mathfrak{S}$, we define its $p$-fold self-composition as $p$-fold
augmented system $\mathfrak{S}^p=(X^p,W^p,f^p)$ where $X^p$ and $W^p$ are $p$-ary
Cartesian powers of $X$, and $W$, respectively, and $f^p: X^p \times W^p \to
X^p$ is equivalent to the $p$-ary Cartesian power of $f$, i.e. 
$f^p: (X \times W)^p \to X^p$ by using the $\zip$ function. 
We use these two types interchangeably. 
We use $\tilde{x}$ and $\tilde{w}$ to denote the state and exogenous input 
of the augmented system $\mathfrak{S}^p$, respectively. 
Similarly, we write $\tilde{\bfm{x}}_{(\tilde{x}_0, \tilde{\nu} )}$ for the \emph{run} of $\mathfrak{S}^p$ starting from an initial state $\tilde{x}_0 \in X^p$  and under exogenous input sequence $\tilde{\nu}: \N \to W^p$. 

\subsection{Barrier Certificates}
A function $\B: X \rightarrow \R$ is a \emph{barrier certificate} for a \dtCS 
$\mathfrak{S} = (X,W,f)$ from the set $A \subseteq X$ to the set $B \subseteq X$, if:
\begin{align}
  \B(x) \leq 0, \quad \quad & \text{ for all } x \in A, \label{eq:bar11}\\
  \B(x) > 0 ,\quad \quad & \text{ for all } x \in B, \label{eq:bar12} 
\end{align}
and for all $x \in X$ and for all $w \in  W$:
\begin{align}
  \B(f(x, w))-\B(x) \leq 0. \label{eq:bar13}
\end{align}
The existence of a barrier certificate is a sufficient condition to
guarantee that if the system $\mathfrak{S}$ ever visits a state from the set $A$, it will 
never visit a state from the set $B$ in the future. 
If the set $A$ characterizes initial states and the set $B$ characterizes the set of bad or 
undesirable states, then the existence of barrier certificates from $A$ to $B$ can guarantee safety, see~\cite{prajna_safety_2004}.  

\subsection{LTL Specifications}

Consider a set of atomic propositions $\AP$ relevant to the underlying system and
the alphabet $\Sigma = 2^{\AP}$ characterized by the subsets of these
propositions. 
We refer to an infinite sequence ($\omega$-sequence) of letters from $\Sigma$ as an infinite trace. We write $\Sigma^\omega$ for the set of all infinite traces over $\Sigma$.

\noindent\textbf{Syntax.} An LTL formula over $\AP$ can be built from the following production rules:
\begin{eqnarray*}
\psi &::=&  a \quad|\quad \neg \psi\quad|\quad \psi \lor \psi \quad|\quad
\nextt \psi \quad|\quad \psi \until \psi,  
\end{eqnarray*}
where $a \in \AP$,  $\nextt$ and $\until$ are the \textsf{next} and \textsf{until} operators, respectively. Other popular temporal operators such as \textsf{globally} ($\always$), \textsf{eventually}
($\eventually$) and \textsf{release} ($\releases$) can be derived from these minimal ones in a
standard manner. 

\vspace{0.5em}

\begin{itemize}
    \item $\psi=a$ and $a \in \bm{\sigma}(0)$,
    \item $\psi= \neg \psi$ and $\bm{\sigma} \not\models \psi$, 
    \item $\psi = \psi_1 \vee \psi_2$ and $\bm{\sigma} \models \psi_1$ or $\bm{\sigma} \models \psi_2$, 
    \item $\psi = \nextt \psi$ and $\bm{\sigma}[1,\infty] \models \psi$, 
    \item $\psi = \psi_1 \until \psi_2$ and $\bm{\sigma}[i,\infty] \models \psi_2$ for some $i\geq 0$ and for all $0 \leq j < i$, we have that $\bm{\sigma}[j,\infty] \models \psi_1$.
\end{itemize}
We refer the interested readers to \cite{baier2008principles} for more details on syntax and semantics of LTL properties. The LTL specifications can only express trace properties, \textit{i.e.}, properties of individual execution traces. However, they cannot specify properties over sets of execution traces, which is essential for many relevant security specifications.

\subsection{HyperLTL Specifications}
\label{subsec:hyperLTL}

HyperLTL, unlike LTL which implicitly considers only a single
trace at a time, can relate multiple traces simultaneously through
the use of existential and universal quantifiers.

\noindent\textbf{Syntax.} We consider HyperLTL with syntax:

\begin{eqnarray*}
\phi &::=& \exists \pi. \phi \quad | \quad \forall \pi. \phi \quad|\quad  \psi \\
\psi &::=& a_\pi \quad\quad|\quad \neg \psi\quad\quad|\quad \psi \lor \psi \quad|\quad
\nextt \psi \quad|\quad \psi \until \psi.
\end{eqnarray*}
The key distinction over LTL formulae is the introduction of trace quantifiers
$\exists$ and $\forall$. The quantifier $\exists \pi$ stands for ``for some
trace $\pi$" while the quantifier $\forall \pi$  stands for ``for all traces
$\pi$", respectively. 
The variable $\psi$ generates standard LTL formulae with the exception that
atomic propositions can refer to distinct trace variables.
Hence, for every proposition $a \in \AP$ and trace variable $\pi$, we
use $a_\pi$ to express that proposition $a$ is referring to the trace $\pi$. 
A trace variable occurs free in a HyperLTL formula, if it is not
bounded by any trace quantifier.
A HyperLTL formula with no free variable is called closed. 

\noindent\textbf{Semantics.} 
Since HyperLTL formulae express the properties of multiple trace variables, one
requires to assign these trace variables to specific traces for reasoning 
about the satisfaction of the formula. Let $\mathcal{V}=\{\pi_1,\pi_2,\ldots\}$ be an infinite set of trace variables. The semantics of a HyperLTL formula $\psi$ is defined over a set $T$ of traces
and a trace valuation function $\Pi: \mathcal{V} \to \Sigma^{\omega}$ that maps
all the free trace variables occurring in the formula $\psi$ to traces in the
set $\Sigma^{\omega}$. 
We use $\Pi[ \pi \to \bm{\sigma}]$ to express the trace valuation function
$\Pi'$ that agrees with $\Pi$ for all trace variables except $\pi$ and
$\Pi'(\pi) = \bm{\sigma}$.
We define the trace valuation suffix $\Pi[i,\infty]$ as $\pi \mapsto
\Pi(\pi)[i, \infty]$, i.e.  $\Pi[i,\infty]$ maps $\pi$ to the $i$-suffix of the
trace mapped to $\pi$ by $\Pi$. 

We say that a HyperLTL formula $\psi$ is valid over a given set $T$ of traces and
trace valuation function $\Pi: \mathcal{V} \to \Sigma^{\omega}$, and we write
$\Pi \models_{T} \phi$ if one of the following holds:
\begin{itemize}
\item
  $\phi = \exists \pi. \psi$ and there is $\bm{\sigma} \in T$ such
  that $\Pi[\pi \to \bm{\sigma}] \models_T \psi$,
\item
  $\phi = \forall \pi. \psi$ and for all $\bm{\sigma} \in T$, we have
  $\Pi[\pi\to \bm{\sigma}] \models_T \psi$,
\item
  $\phi = a_\pi$ and $a \in \Pi(\pi)(0)$, 
\item
  $\phi = \neg \phi$ and $\Pi \not\models_T \phi$,
\item
  $\phi =  \psi_1 \lor \psi_2$ and $\Pi \models_T \psi_1$ or $\Pi \models_T \psi_2$,
\item
  $\phi = \nextt \psi$ and $\Pi[1,\infty] \models_T \psi$, 
\item
  $\phi = \psi_1 \until \psi_2$ and there is $i \geq 0$ such that
  $\Pi[i,\infty] \models_T \psi_2$ and for all $0 \leq j < i$, we have that
  $\Pi[j,\infty] \models_T \psi_1$.
\end{itemize}
A closed HyperLTL formula $\phi$ is considered to be satisfied by a set of
traces $T$, and we write $T \models \phi$ if the empty trace assignment
satisfies the formula, \textit{i.e.}, $\emptyset \models_T \phi$. We refer the interested readers to \cite{clarkson_temporal_2014} for more details on syntax and semantics of HyperLTL properties. 

\subsection{B\"uchi Automata}
\label{sec:automata}
A B\"uchi automaton is a tuple ${\mathcal A} =
(\Sigma, Q, q_0, \Delta, F)$, where $\Sigma$ is a finite set of \emph{alphabet},
$Q$ is a finite set of \emph{states}, $q_0 \in Q$ is the \emph{initial state}, 
$\Delta \subseteq Q  \times \Sigma \times Q$ is the transition relation and $F \subseteq Q$ is the set of \emph{accepting states}. 
We say that a B\"uchi automaton is \emph{deterministic} if $\Delta$ can be represented as a function that maps a state and alphabet pair to exactly one state, i.e., $\Delta: Q \times \Sigma \to Q$, and nondeterministic otherwise. We abbreviate deterministic and nondeterministic B\"uchi automata as DBA and NBA respectively.
A \emph{run} $r$ of ${\mathcal A}$ on $\textsc{w} \in \Sigma^\omega$ is an
$\omega$-word $r_0, \textsc{w}_0, r_1, \textsc{w}_1, \ldots$ in
$(Q \times \Sigma)^\omega$ such that $r_0 = q_0$ and, for $i > 0$, $(r_{i-1},
\textsc{w}_{i-1},r_i)\in \Delta$.
We write $\infi(r)$ for the set
of states that appear infinitely often in the run $r$. A word $\textsc{w}$ is said to be accepted by the NBA $\mathcal A$, if, for the corresponding run $r$, we have $\infi(r) \cap F \neq \emptyset$. The \emph{language}, $\mathcal{L}_{\mathcal{A}}$, of any automaton ${\mathcal A}$ (or, \emph{recognized} by
${\mathcal A}$) is the subset of words in $\Sigma^\omega$ that have accepting runs
in ${\mathcal A}$.
A language is $\omega$-\emph{regular} if it is accepted by an NBA.

It is well known that for any LTL specification, one may compile a corresponding NBA~\cite{LTL_Buchi_Vardi_wolper} by using LTL to B\"uchi construction techniques~\cite{LTL_to_Buchi}. 
However, in the case of HyperLTL, the presence of quantification of traces in the specification prevents an immediate compilation to an automaton. Nevertheless, given a HyperLTL specification $\phi=\mu_1 \pi_1 \ldots \mu_p \pi_p \psi$ defined over a set of
atomic propositions $\AP$, where for all $ 1 \leq i \leq p$, $\mu_i \in \{\exists, \forall\}$, one may construct an NBA $A_{\psi}$ corresponding to the quantifier-free LTL formula $\psi$ over the alphabet $\Sigma^p$.

In order to accommodate for the quantification of traces, a word for this automaton is a $p$-tuple of individual
traces, denoted by
$\tilde{\bm{\sigma}}=(\bm{\sigma}_1,\ldots,\bm{\sigma}_p$), i.e., $\unzip(\tilde{\bm{\sigma}}) \in \mathcal{L}_{\mathcal{A}_\psi}$ if $(\pi_1 \mapsto \bm{\sigma}_1, \ldots, \pi_p \mapsto \bm{\sigma}_p)$ 
is an accepting run of $\mathcal{A}_{\psi}$. 

\subsection{HyperLTL Verification Problem}
Let $\mathfrak{S}=(X,W,f)$ be a \dtCS and $L: X \rightarrow \Sigma$ be a
measurable function labeling states of \dtCS $\mathfrak{S}$ with letters of the alphabet.
We can extend the labeling function from states to runs of $\Ss$ in a
straightforward fashion: for every infinite state run
$\bfm{x}=(\bfm{x}(0),\bfm{x}(1),\ldots)$ of $\Ss$, the function $L(\bfm{x})$
gives the corresponding trace of the system $(\sigma_0, \sigma_1,\ldots) \in
(\Sigma)^\omega$, where $\sigma_i = L(\bfm x(i))$, for all $i \in \mathbb{N}$. 
Let $T(\Ss, L)$ be the set of all traces of $\Ss$.
We now define the key problem of interest. 
\begin{problem}[HyperLTL Verification] \label{probstat}
Given a discrete-time uncertain dynamical system $\mathfrak{S} = (X,W,f)$, a labeling function $L: X \rightarrow
\Sigma$, and a HyperLTL specification $\phi$, the HyperLTL verification problem $\mathfrak{P} = (\mathfrak{S}, L, \phi)$ is to 
decide whether $T(\Ss, L) \models \phi$.
\end{problem}

While model checking of finite-state systems against HyperLTL specifications is 
decidable~\cite{Finkbeiner2015AlgorithmsFM}, the verification problem stated above is in general 
undecidable for continuous state-space systems considered in this paper. 
It follows readily from the fact that even simple reachability is undecidable for simple continuous state-space 
dynamical systems~\cite{ASARIN199535}.  
Our approach provides a sound procedure for Problem \ref{probstat}.

%% file: problem.tex
 \begin{figure*}[t]
 \centering
 \vspace{-0.8em}
     \includegraphics[scale=0.9]{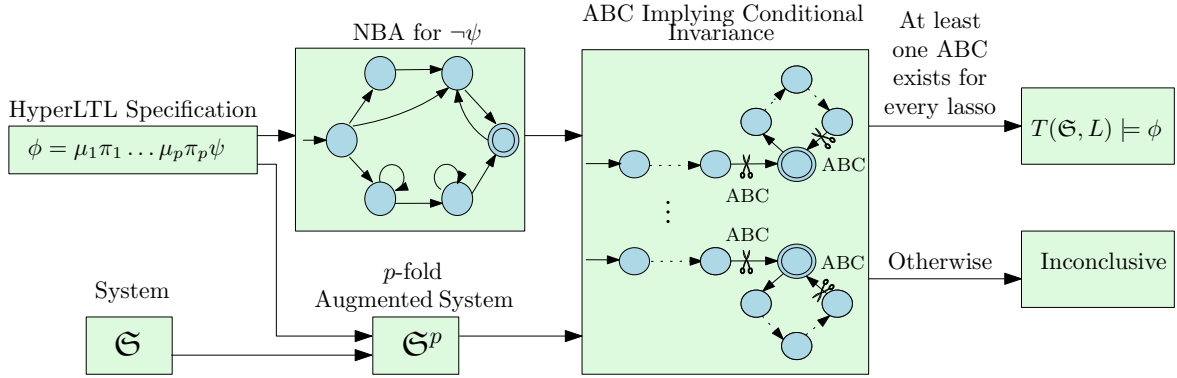}
     \caption{A schematic block diagram illustrating the verification procedure.}
     \label{fig:block_diag}
 \end{figure*}

Verification of HyperLTL formulae in the context of finite systems has been well studied~\cite{Finkbeiner2015AlgorithmsFM,clarkson_temporal_2014}. The verification procedure is based on automata-theoretic model-checking, where quantifier-free fragments of the desired HyperLTL formulae are compiled into $\omega$-automata, and trace quantification is handled by appropriately composing these automata with the underlying Kripke Structure. The interleavings of automata and Kripke products lead to an automata whose language emptiness decides the satisfaction of the specification. Unfortunately, this approach cannot be extended to continuous state-space systems by simply using abstraction-based techniques as system relations (e.g., simulation relations) may not preserve hyperproperties~\cite{opacity_simu}.

To verify the specification $\phi$ against \dtCS $\mathfrak{S}$, we compile the negation of the HyperLTL formula  into  an
implicitly quantified B\"uchi automata, and discharge the unsatisfiability of
the specification to synthesizing appropriate barrier certificates. 
Our verification procedure is depicted in Figure \ref{fig:block_diag}.
Here, given a HyperLTL specification $\phi$ and system $\mathfrak{S}$, we construct the $p$-fold augmented system $\mathfrak{S}^p$, and the NBA for $\neg \psi$. We then find an augmented barrier certificate (ABC) that acts as a proof certificate of conditional invariance (cf. Definition \ref{def:cond_inv}) for some transitions along every lasso. This acts as a ``scissor'' and allows us to conclude that the system $\mathfrak{S}$ satisfies the specification $\phi$. We note that our approach is not complete in that if it cannot find an ABC for at least one transition along every lasso then we cannot conclude that the system does not satisfy $\phi$.
The rest of the section introduces the idea of augmented barrier certificates and provides an automata-theoretic sound verification approach for Problem~\ref{probstat}.

\subsection{HyperLTL Evaluation Game Semantics}
We provide a game semantics to the HyperLTL verification problem $\mathfrak{P} = (\mathfrak{S}, L, \phi)$ as a two-player 
{\it stage-based evaluation game} played between two players, Eloise ($\exists$) and Abelard ($\forall$),  where 
Eloise takes the role of a verifier and her goal is to prove that $T(\Ss, L) \models \phi$, while 
the goal of Abelard (the spoiler) is the opposite.  
Given a HyperLTL formula $\mu_1 \pi_1 \ldots \mu_p \pi_p \psi$, we say 
that Eloise controls the quantifier $\mu_i$ if $\mu_i = \exists$, otherwise we say that Abelard controls the quantifier.  
The game continues in stages.
In the first stage, the game begins with a token in the initial position $(\mu_1 \pi_1 \mu_2 \pi_2 \ldots \mu_p \pi_p \psi, \Pi=\emptyset)$ \cite{Goranko_Game_Theoretic_Semantics} and the player    
controlling the left-most quantifier $\mu_1$ chooses a trace $\bm{\sigma_1}$ from $T(\Ss, L)$ and moves the token 
to the next position $(\mu_2 \pi_2 \ldots \mu_p \pi_p \psi, \Pi=\set{\pi_1 \to \bm{\sigma_1}})$.  

The game from the next position continues in a similar fashion until we reach a position with a quantifier-free HyperLTL formula. 
We call such positions {\it terminal}.
We say that a terminal position $(\psi, \Pi=\{\pi_1 \to \bm{\sigma_1},\ldots,$ $ \pi_p \to \bm{\sigma_p}\})$ 
is winning for Eloise if $\zip(\bm{\sigma_1},\ldots,\bm{\sigma_p}) \models \psi$ with the standard LTL semantics for the formula $\psi$.
We say that Eloise wins the multi-stage evaluation game if she has a way to choose her moves such that no matter how Abelard chooses his moves
the game ends in a winning position for Eloise; otherwise Abelard wins the game.
Notice that at every step, both players have access to complete infinite traces that have been chosen by players in earlier 
positions as a part of position description. 
It is trivial to see that Eloise wins the evaluation game if and only if $\Pi \models_{T} \phi$.

We consider another version of evaluation games that we dub {\it turn-based evaluation games}.
These games are played on the $p$-fold self-composition $\mathfrak{S}^p=(X^p,W^p,f^p)$ of the system $\mathfrak{S}$.
These games start with a token in some initial configuration $(x_1^0, x_2^0, \ldots, x_p^0) \in X^p$ and at every round first 
the player controlling $\mu_1$ chooses an action $w_1\in W$, followed by the player controlling $\mu_2$, and so on. 
In this way, the players form a set of joint actions $(w_1, w_2, \ldots, w_p)\in W^p$ and the token is moved to a state 
$f^p((x_1^0, x_2^0, \ldots, x_p^0), (w_1, w_2, \ldots, w_p))$.
The game continues in this fashion indefinitely and the players thus form an infinite run $\tilde{\bfm{x}}$  
of $\mathfrak{S}^p=(X^p,W^p,f^p)$. 
To relate the augmented system $\mathfrak{S}^p$ with letters in $\Sigma^p$, we extend the definition of the labeling function to the augmented system domain by using $L^p: X^p \rightarrow \Sigma^p$ to map the states in the augmented state set to the alphabet $\Sigma^p$.
We say that the trace $\tilde{\bfm{x}}$ is winning for Eloise if $L^p(\tilde{\bfm{x}}) \models \psi$.
We say that Eloise has a winning strategy in the turn-based evaluation game for $\mathfrak{P} = (\mathfrak{S}, L, \phi)$ if she 
can select her moves in such a way that no matter how Abelard chooses his moves (including using an arbitrary look-ahead), the resulting trace is winning for Eloise.
Moreover, we say that Eloise has a positional winning strategy 
if to select actions in a given round her choice depends only on the current states and choices resolved before her turn 
for the other quantifiers in the current round.  

\begin{lemma}
 \label{lem:weaken}
If Eloise has a positional winning strategy in a turn-based evaluation game for  $\mathfrak{P}$ 
then she has a winning strategy in the stage-based evaluation game. 
\end{lemma}
\subsection{Augmented Barrier Certificates}
\label{sec:barrier}
\input{barrier_new}

\subsection{ABCs as ``scissors''}
\label{sec:scissor}
\input{scissor}

%% file: barrier_new.tex
We reduce the search for a positional strategy for Eloise in a turn-based evaluation game 
to the search for barrier-functions like certificates that we call augmented barrier 
certificates (ABCs).
Just like barrier certificates provide a proof that separates two sets over $X$ for arbitrary traces of the system, ABCs provide a proof that separate two sets over $X^p$ for appropriately chosen traces 
by players.
To tie in the notion of ABCs with HyperLTL properties, we present a special class of properties that we call 
{\it conditional invariance} properties  that generalize the notion of invariance.
 
\begin{definition}[Conditional Invariance (CI)]
\label{def:cond_inv}
We say that a HyperLTL formula $\chi {=} \mu_1 \pi_1 \ldots \mu_p \pi_p \xi$ is a conditional 
invariance (CI) if $\xi$ is of the form $\always( s_A \rightarrow \always( \neg s_B)) $ 
where $s_A, s_B$ are Boolean combination of atomic propositions. 
\end{definition}

\begin{definition}[Augmented Barrier Certificates (ABCs)]
\label{def:ABC}
Consider a CI $\chi = \mu_1 \pi_1 \ldots \mu_p \pi_p \always( s_A \rightarrow \always( \neg s_B))$ and
the sets  ${L^p}^{-1}(s_A) = A \subseteq X^p$ and ${L^p}^{-1}(s_B) = B \subseteq X^p$. 
We say that $\B: X^p \rightarrow \R$ is an \emph{augmented barrier certificate (ABC)} 
for a \dtCS $\mathfrak{S} = (X,W,f)$ and property $\chi$ from the set $A \subseteq X^p$ to set $B \subseteq X^p$ if
\begin{align}
  \B(\tilde{x}) \leq 0, \quad \quad &\text{ for all } \tilde{x} \in A, \label{eq:bar1}\\
  \B(\tilde{x}) > 0 ,\quad \quad &\text{ for all } \tilde{x} \in B, \label{eq:bar2}
   \vspace{-1em}
\end{align}
and  $\forall x_1 \in X,$ $\mu_1 w_1 \in W,$ $\forall x_2 \in X,$ $\mu_2 w_2 \in W, \ldots,$ $\forall x_p \in X,$ $\mu_p w_p \in W$ one has:
\begin{align}
  \B(f^p(\tilde{x},\tilde{w}))-\B(\tilde{x}) \leq 0, \label{eq:bar3}
\end{align}
where $\tilde{x} = (x_1, \ldots, x_p)$ and $\tilde{w} = (w_1, \ldots, w_p)$. 
\end{definition}

\begin{remark}
Note that all the states' components $x_i$ of the augmented system $\mathfrak{S}^p$ in condition \eqref{eq:bar3} are quantified universally, while their corresponding exogenous inputs' components $w_i$ are quantified according to $\mu_i$. The components $x_i$ cannot be quantified according to $\mu_i$ as it may result in the state runs of the augmented system $\mathfrak{S}^p$ reaching the unsafe region $B$. To see this, consider a component $x_i(t)$ at some time step $t \in \N$ which is quantified by $\mu_i= \exists$. For that component, one may be able to pick a corresponding input component $w_i(t)$ such that the augmented barrier certificate is non-increasing according to condition~\eqref{eq:bar3}. However, since $x_i$ is quantified only existentially, one may fail to ensure the existence of input $w_i(t+1)$ for $x_i(t+1)$ at the next time step such that the augmented barrier certificate is still non-increasing. Due to this, one would fail to ensure that the augmented barrier certificate remains non-increasing at every time step, possibly resulting in safety violations.

\end{remark}

\begin{remark}\label{remarkdecomp}
  For a CI $\chi =  \mu_1 \pi_1 \ldots \mu_p \pi_p \always(s_A {\rightarrow} \always(\neg s_B ) )$, if the set ${L^p}^{-1}(s_A) \cap {L^p}^{-1}(s_B) $ is non-empty, then there exists no ABC satisfying 
  conditions \eqref{eq:bar1}-\eqref{eq:bar3}. This is due to the conflict in the satisfaction of conditions 
  \eqref{eq:bar1} and \eqref{eq:bar2}.
\end{remark}

\begin{lemma}
\label{lem:ABCLogic}
The existence of an ABC for a conditional invariance $\chi$ implies 
that $T(\Ss, L) \models \chi$.
\end{lemma}

%% file: scissor.tex
\begin{algorithm}
\begin{algorithmic}
\Require{$\mathfrak{S},\phi= \mu_1 \pi_1\ldots\mu_p \pi_p \psi, L$ }
\State Construct NBA $\mathcal{A}_{\neg \psi}$ for $\neg \psi$
\State Identify lassos $\mathcal{R} = \{ \bfm{r}_1,\bfm{r}_2,\ldots,\bfm{r}_k \}$ of $\mathcal{A}_{\neg \psi}$
\For {$i\gets1$ to $k$}
\State Identify consecutive transition pairs 
\State $\mathcal{S}_{\mathcal{A}_{\neg \psi}} = \{(s_{A_{i,1}},s_{B_{i,1}}),\ldots,(s_{A_{i,v_i}},s_{B_{i,v_i}}) \}$
\For{$j \gets 1$ to $v_i$}
\State $\chi_{i,j} \gets \mu_1 \pi_1 \ldots \mu_p \pi_p \always (s_{A_{i,j}} \rightarrow \always(\neg s_{B_{i,j}}))$
\EndFor
\EndFor
\State Construct augmented system $\mathfrak{S}^p$
\State Find common ABC $\B$ for $\chi_{i,j}$ for all $i \in \{1,\ldots,k\}$ and some $j \in \{1,\ldots,v_i\}$
\If{$\B$ exists}
\State \textbf{return} $T(\mathfrak{S},L) \models \phi$
\Else
\State \textbf{return} Inconclusive
\EndIf
\end{algorithmic}
\caption{Algorithm for verification of HyperLTL formulae}
\label{alg:procedure}
\end{algorithm}

To extend CI guarantees obtained via ABCs to arbitrary HyperLTL specifications, we first construct an NBA $\mathcal{A}_{\neg \psi}=(\Sigma^{p},Q,q_0,\Delta,F)$ corresponding to $\neg \psi$. 
Then, we employ ABCs as \emph{scissors} disallowing transitions to the accepting states of $\mathcal{A}_{\neg \psi}$. To do so, we first present the following lemmas to guarantee disjunction or conjunction over a set of CIs.

\begin{lemma}[ABCs for disjunction of CIs]
    \label{lem:DisjunctABC}
    Given a set of CIs  $\{\chi_1, \ldots, \chi_k\}$, the existence of an ABC $\B_j$ for some CI $\chi_j$, where $\chi_j = \mu_1 \pi_1 \ldots \mu_p \pi_p \always(s_{A_j} \rightarrow \always(\neg s_{B_j}))$ for $1 \leq j \leq k$, implies that $T(\mathfrak{S},L) \models \chi$, where 
    \[ 
        \chi = \mu_1 \pi_1 \ldots \mu_p \pi_p  \underset{1 \leq j \leq k}{\bigvee}\always (s_{A_j} \rightarrow \always(\neg s_{B_j})).
    \]
\end{lemma}
\vspace{0.5em}

We say that a function $\B$ is a common ABC for a set of CIs $\{\chi_1, \ldots, \chi_k\}$ if $\B$ is an ABC for all of the CIs.    \begin{lemma}[ABCs for conjunction]
    \label{lem:Common_ABC}
    The existence of a common ABC for a set of CIs $\{\chi_1, \chi_2, \ldots, \chi_k\}$, where $\chi_i = \mu_1 \pi_1 \ldots \mu_p \pi_p \always(s_{A_i} \rightarrow \always(\neg s_{B_i}) )$, for $ 1 \leq i \leq k$, implies that $T(\mathfrak{S},L) \models \chi$, where  
\[\chi = \mu_1 \pi_1 \ldots \mu_p \pi_p  \underset{1 \leq i \leq k}{\bigwedge}\always (s_{A_i} \rightarrow \always(\neg s_{B_i})).\]
\end{lemma}
\vspace{0.5em}

\begin{lemma}[ABCs for conjunctive normal forms]
    \label{lem:ConjunctDisjunctABC}
        Given a set of sets of conditional invariances 
        \[
        \{ \{\chi_{1,1}, \ldots,  \chi_{1,v_{1}}\}, \{ \chi_{2,1}, \ldots, \chi_{2,v_{2}} \}, \ldots \{ \chi_{k,1}, \ldots, \chi_{k,v_{k}} \} \},
        \]
    where $\chi_{i,j} = \mu_1 \pi_1 \ldots \mu_p \pi_p \always(s_{A_{i,j}} \rightarrow \always(\neg s_{B_{i,j}}) )$, the existence of a common ABC for $\chi_{i,j}$ for every $1 \leq i \leq k$ and some $1\leq j\leq v_i$  implies that $T(\mathfrak{S},L) \models \chi$, where 
        \[\chi = \mu_1 \pi_1 \ldots \mu_p \pi_p  \underset{(1 \leq i \leq k) }{\bigwedge } \underset{ (1 \leq j \leq v_{i}) }{\bigvee }\always (s_{A_{i,j}} \rightarrow \always(\neg s_{B_{i,j}})).\]
    \end{lemma}
    \vspace{0.5em}

To find the solution to Problem~\ref{probstat}, we consider the NBA $\mathcal{A}_{\neg \psi}$ corresponding to $\neg \psi$, obtained from the HyperLTL specification $\phi= \mu_1 \pi_1 \ldots \mu_p \pi_p \psi$. Then, we decompose $\mathcal{A}_{\neg \psi}$ to consecutive transition pairs along its accepting lassos (or simply lassos). Such a lasso consists of a simple path from the initial state to some accepting state, followed by a simple cycle on the accepting state. Note that the number of lassos in $\mathcal{A}_{\neg \psi}$ is finite since the NBA has finitely many edges that lead to finitely many simple paths to an accepting state and simple cycles over an accepting state. Any two consecutive edges along these lassos constitute a transition pair which corresponds to a CI specification. Utilizing Lemma~\ref{lem:ConjunctDisjunctABC}, the following theorem characterizes a condition to solve Problem~\ref{probstat}. 
\begin{theorem}
\label{thm:main_result}
Given a HyperLTL specification $\phi = \mu_1 \pi_1 \ldots \mu_p \pi_p \psi$, the existence of a common ABC $\B$ for some consecutive transition pair along every lasso of $A_{\neg \psi}$ guarantees that $T(\mathfrak{S},L) \models \phi$.
\end{theorem}

To decompose the NBA $\mathcal{A}_{\neg \psi}$ into consecutive transition pairs $(s_{A}, s_{B})$, one can view $\mathcal{A}_{\neg \psi}$ as a graph and utilize variants of depth-first search algorithms~\cite{russell_artificial_2009}. This decomposition of the NBA into a collection of transition pairs was presented in~\cite{wongpiromsarn_automata_2016,jagtap_temporal_2018} but in the context of verification for LTL specifications. Algorithm~\ref{alg:procedure} demonstrates a sound verification procedure used in this paper. We refer the interested readers to~\cite[Section 4]{jagtap_temporal_2018} for more details on the decomposition procedure.
\begin{remark}
\label{remark:Complexity}
We note that the problem of finding the collection of consecutive transition pairs (thus conditional invariances), one from each lasso, which admit a common ABC is  intractable. 
To show this, we first assume that we are given an oracle that determines whether a collection of consecutive transition pairs admits a common ABC. We then consider a relaxed version of this problem as follows. We assume that for any state $r$ in $\mathcal{A}_{\neg \psi}$ with some incoming edge labeled $s_A$ and outgoing edges $s_{B_1}, \ldots, s_{B_r}$, if there exists an ABC $\B$ for the pair $(s_A,s_{B_j})$ for some $1 \leq j \leq r$, then the function $\B$ acts as an ABC for every pair $(s_A,s_{B_j})$ for all $1 \leq j \leq r$. Then, the problem of finding a suitable collection of transition pairs is reduced to finding a collection of edges such that their removal causes the accepting states to not be reachable from the initial state. This corresponds to a cut~\cite{algorithms_CLRS} that partitions the accepting states from the initial state. To determine whether a cut allows for a common ABC, we must make use of the oracle, and in the worst case, we need to enumerate all possible cuts in $A_{\neg \psi}$. Since, the number of possible cuts is exponential in the number of edges of $A_{\neg \psi}$~\cite{algorithms_CLRS}, the problem is clearly intractable. 
\end{remark}

The requirement of a common ABC for a collection of consecutive transition pairs is necessary to provide guarantees via Theorem~\ref{thm:main_result}. This is due to the fact that, in condition~\eqref{eq:bar3} of Definition~\ref{def:ABC}, existential quantifiers may precede the universal quantifiers depending on the HyperLTL specification. In such cases, Eloise does not have access to the full-state information of the augmented system and the choices made by Abelard in the turn-based game. However, Abelard has access to the states as well as Eloise's choices. Then, different ABCs for different transition pairs would imply that for each transition pair, Eloise picks a different strategy. This may lead to conflicts. For example, let us assume that the HyperLTL formula is of the form $\phi = \exists \pi_1 \forall \pi_2 \psi $, and consider two conditional invariances corresponding to pairs $(s_{A_1},s_{B_1})$ and $(s_{A_2},s_{B_2})$. The first component of the state of the augmented system is controlled by Eloise, and the second one by Abelard. Due to a lack of full-state information of the augmented system for Eloise, she is only able to observe the label of the first component, and therefore may be unable to differentiate between $s_{A_1}$ and $s_{A_2}$. Thus, having two different strategies corresponding to each of these pairs may result in ambiguity for Eloise. Moreover, picking the first strategy corresponding to $(s_{A_1},s_{B_1})$ at state $s_{A_2}$ could lead to Abelard choosing an input that violates the second conditional invariance corresponding to $(s_{A_2},s_{B_2})$, and  vice-versa as Abelard selects a trace after Eloise selects her trace. This results in violation of the original specification.
Unfortunately, even though a common ABC is necessary to provide verification guarantees, its existence may be difficult to find.

However, in specifications where Eloise has access to full state information and all of Abelard's choices, the requirement of a common ABC may be relaxed. This is especially true for specifications in the $\forall^{*}\exists^{*}$ fragment, where all the universal quantifiers precede the existential ones. In fact, the $\forall^{*}\exists^{*}$ fragment holds great importance as it comprises of many relevant security properties. For example, a variant of the noninterference property~\cite{noninterference2} requires that, for all traces, the low-security variables should not see any difference in observation when high-security variables are changed and replaced by dummy variables. This can be expressed by the HyperLTL specification  
\[
\forall \pi_1 \exists \pi_2 (\always h_{\pi_2}) \wedge \underset{l \in LS}{\bigwedge} l_{\pi_1} \leftrightarrow l_{\pi_2},
\]
where $h_{\pi_2}$ implies that the high security variables  in $\pi_2$ are all set to a dummy variable $h$ that is always true
, and $LS \in \AP$ denotes the set of low security variables. Similarly, initial-state opacity specification~\cite{opacity_simu}
is also in the $\forall^{*}\exists^{*}$ fragment (cf. case study). Considering the importance of this fragment, we now provide a separate algorithm to allow for multiple ABCs for different lassos under some conditions.
\vspace{-1em}
\subsection{Algorithm for $\forall^{*}\exists^{*}$-fragment of HyperLTL}
\label{subsec:forallexists}
From the above discussion, it can be understood that specifications in the $\forall^{*}\exists^{*}$ fragment enable the relaxation of common ABC requirement and allow for different ABCs in different lassos. In particular, Eloise can take advantage of the full state information of the augmented system available to her as well as the knowledge of Abelard's choices to use different ABCs for different consecutive transition pairs in every lasso. However, to do so, one must take the structure of the automata $\mathcal{A}_{\neg \psi}$ into consideration, as in the presence of states with two or more outgoing edges, there may be an ambiguity for Eloise in selecting strategies. Moreover, in the presence of nondeterminism in the automaton, Eloise may fail to select strategies due to lack of information on the history of visited states. 
These challenges are demonstrated in the following examples.

\begin{example}[States with a fork] \label{ex:diffABC}
In this example, we show the issue of utilizing multiple ABCs in the presence of a state with multiple outgoing edges. Consider the NBA $\mathcal{A}_{\neg \psi}$ shown in Figure~\ref{fig:NBA_AE_eg1} constructed from a set of atomic propositions $\AP=\{a,b,c,d\}$ corresponding to some HyperLTL specification $\phi=\forall \pi_1 \ldots \forall \pi_l \exists \pi_{l+1} \ldots \exists \pi_p \psi$. 
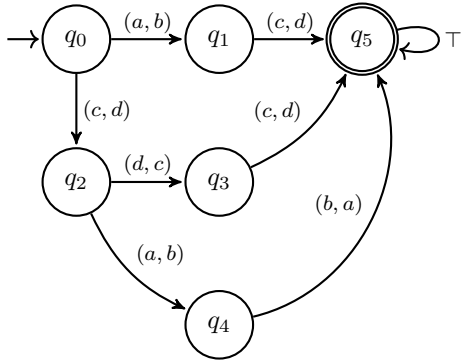
\begin{figure}[h]
    \centering
   \begin{tikzpicture}[node distance=1.9cm,thick,
 el/.style = {inner sep=2pt, align=left},
 every label/.append style = {font=\tiny},
 every  edge/.append style = {draw, -stealth', shorten > = 1pt,
                              font=\footnotesize, inner sep=2pt, auto}]
    
     \node[initial, state, initial text=, ] (0) {$q_0$};
     \node[state, right of = 0, ] (1) {$q_1$};
     \node[state, accepting, right of = 1] (5) {$q_5$};
     \node[state, below of = 0] (2) {$q_2$};
     \node[state, right of = 2] (3) {$q_3$};
     \node[state, below of = 3] (4) {$q_4$};
     \path [->]
     (0) edge node{$(a,b)$} (1) 
     (0) edge node{$(c,d)$} (2)
     (1) edge node{$(c,d)$} (5)
     (3) edge [bend right = 20] node{$(c,d)$} (5)
     (2) edge node{$(d,c)$} (3) 
     (2) edge [bend right = 20] node{$(a,b)$} (4) 
     (4) edge [bend right = 50] node{$(b,a)$} (5) 
     (5) edge [loop right] node{$\top$} (5);
     \end{tikzpicture}
    \caption{NBA $A_{\neg \psi}$ for Example \ref{ex:diffABC}}
    \label{fig:NBA_AE_eg1}
\end{figure}

From $\mathcal{A}_{\neg \psi}$, we can identify $k=3$ lassos as 
\begin{equation*}
    \mathcal{R}=\{ \bfm{r}_1=(q_0,q_1,q_5,q_5), \bfm{r}_2=(q_0,q_2,q_3,q_5,q_5), \bfm{r}_3= (q_0,q_2,q_4,q_5,q_5)\}.
\end{equation*}
For every $\bfm{r} \in \mathcal{R}$, we obtain the consecutive transition pairs as
\begin{align*}
    &\mathcal{S}(\bfm{r}_1)= \{((a,b),(c,d)), ((c,d),\top)\}, \\ 
    &\mathcal{S}(\bfm{r}_2)= \{((c,d),(d,c)),((d,c),(c,d)),((c,d),\top)\}, \\
    &\mathcal{S}(\bfm{r}_3)= \{((c,d),(a,b)),((a,b),(b,a)),((b,a),\top)\}.
\end{align*}

Naturally, it is preferable to obtain different ABCs for at least one transition pair in every lasso to guarantee the satisfaction of the specification. However, this might cause problems for lassos $\bfm{r}_2$ and $\bfm{r}_3$, where there are two outgoing edges from a single state $q_2$. This leads to two different transition pairs $((c,d),(d,c))$ and $((c,d),(a,b))$. Having different ABCs for these pairs would result in different winning strategies for Eloise to avoid the sets corresponding to $(d,c)$ and $(a,b)$, from the set corresponding to $(c,d)$, respectively. Choosing the first ABC and its corresponding strategy could lead to the violation of condition \eqref{eq:bar3} for the second ABC and vice versa. However, the existence of a common ABC for both the pairs guarantees that Eloise has a winning strategy to avoid both $(d,c)$ and $(a,b)$ if she encounters a state corresponding to $(c,d)$. Therefore, for this specification, one would require to obtain a common ABC for the pairs $((c,d),(d,c))$ and $((c,d),(a,b))$ from lassos $\bfm{r}_2$ and $\bfm{r}_3$, respectively, and a different ABC may be obtained for the pair $((a,b),(c,d))$ from the lasso $\bfm{r}_1$. However, if such a common ABC cannot be found, one can consider other transition pairs in $\bfm{r}_2$ and $\bfm{r}_3$, and in that case, different ABCs may be used.
\end{example}
\begin{example}[Nondeterminism] \label{ex:diffABC1}

In this example, we show the issue of using multiple ABCs in the presence of nondeterminism in the automaton. Consider the NBA $\mathcal{A}_{\neg \psi}$ shown in Figure~\ref{fig:NBA_AE_eg2} constructed from a set of atomic propositions $\AP=\{a,b,c,d\}$ corresponding to some HyperLTL specification $\phi=\forall \pi_1 \ldots \forall \pi_l \exists \pi_{l+1} \ldots \exists \pi_p \psi$. 

\begin{figure}[h]
    \centering
  \begin{tikzpicture}[node distance=1.9cm,thick,
 el/.style = {inner sep=2pt, align=left},
 every label/.append style = {font=\tiny},
 every  edge/.append style = {draw, -stealth', shorten > = 1pt,
                              font=\footnotesize, inner sep=2pt, auto}]
    
     \node[initial, state, initial text=, ] (0) {$q_0$};
     \node[state, right of = 0, ] (3) {$q_3$};
     \node[state, right of = 3, ] (4) {$q_4$};
     \node[state, accepting, right of = 4] (5) {$q_5$};
     \node[state, above of = 3] (1) {$q_1$};
     \node[state, right of = 1] (2) {$q_2$};
     \path [->]
     (0) edge node{$(a,b)$} (1)
     (0) edge node{$(a,b)$} (3) 
     (1) edge node{$(b,a)$} (2)
     (2) edge node{$(a,c)$} (5)
     (3) edge node{$(b,a)$} (4)
     (4) edge node{$(c,d)$} (2)
     (4) edge node{$(d,c)$} (5) 
     (5) edge [loop right] node{$\top$} (5);
     \end{tikzpicture}
    \caption{NBA $A_{\neg \psi}$ for Example \ref{ex:diffABC1}}
    \label{fig:NBA_AE_eg2}
\end{figure}
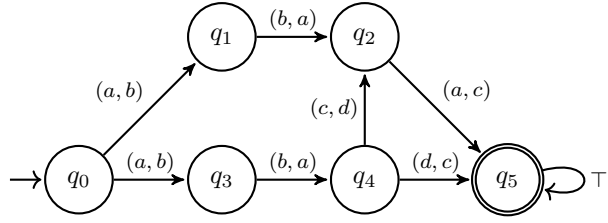

From $\mathcal{A}_{\neg \psi}$, we can identify $k=3$ lassos as 
\begin{equation*}
    \mathcal{R}=\{\bfm{r}_1=(q_0,q_1,q_2,q_5,q_5), \bfm{r}_2=(q_0,q_3,q_4,q_5,q_5), \bfm{r}_3=(q_0,q_3,q_4,q_2,q_5,q_5)\}.
\end{equation*}
For every $\bfm{r} \in \mathcal{R}$, we obtain the consecutive transition pairs as
\begin{align*}
    &\mathcal{S}(\bfm{r}_1)= \{((a,b),(b,a)), ((b,a),(a,c)), ((a,c),\top)\}, \\ 
    &\mathcal{S}(\bfm{r}_2)= \{((a,b),(b,a)),((b,a),(d,c)),((d,c),\top)\}, \\
    &\mathcal{S}(\bfm{r}_2)= \{((a,b),(b,a)),((b,a),(c,d)),((c,d),(a,c)),((a,c),\top)\}.
\end{align*}
Consider lassos $\bfm{r}_1$, $\bfm{r}_2$ and $\bfm{r}_3$, where there is a nondeterministic transition from the initial state $q_0$ to the states $q_1$ and $q_3$ under the label $(a,b)$. Ideally, a single ABC for the pair $((a,b),(b,a))$ would effectively disallow the transitions in all the lassos $\bfm{r}_1$, $\bfm{r}_2$ and $\bfm{r}_3$. However, the problem arises when such an ABC cannot be found. In order to guarantee the satisfaction of the specification, other transition pairs in the lassos must be disallowed. Now, in $\bfm{r}_1$, there is a transition from $(b,a)$ to $(a,c)$, while in $\bfm{r}_2$ and $\bfm{r}_3$, there are two transitions from $(b,a)$ to $(d,c)$ and $(c,d)$, respectively. At any point in time, Eloise cannot uniquely determine the history of the states visited in $\mathcal{A}_{\neg \psi}$. As a result, after a nondeterministic transition from $(a,b)$ to $(b,a)$, Eloise has no way of knowing whether to block further transitions from $(b,a)$ to $(a,c)$, or from $(b,a)$ to $(d,c)$ and $(c,d)$. Therefore, the approach of using different ABCs fails in the presence of nondeterminism.
\end{example}

Unfortunately, problems arising due to nondeterminism cannot be directly resolved. Therefore, to circumvent this issue, we instead consider the automaton $\mathcal{A}_{\neg \psi}$ to be deterministic. 
In particular, for our exposition we focus on the case where $\mathcal{A}_{\neg \psi}$ is a deterministic B\"uchi automaton (DBA)  and later show how to extend this approach for nondeterministic B\"uchi automata (NBA).

The general verification procedure for determining whether Eloise has a strategy to ensure that the acceptance condition of $\mathcal{A}_{\neg \psi}$ is violated for specifications in the $\forall^{*}\exists^{*}$ fragment is provided in Algorithms~~\ref{alg:AEprocedure} and \ref{alg:ABC_FIND}. Having a DBA $\mathcal{A}_{\neg \psi}=(\Sigma^p,Q,q_0,\Delta,F)$, we first identify all the lassos $\mathcal{R}=\{\bfm{r}_1,\ldots,\bfm{r}_k\}$ that reach and cycle on some state in $F$. We then prune $\mathcal{A}_{\neg \psi}$ and remove any states that are not in the lassos and all transitions to and from such states. Then, beginning from the initial state, for any label $s_{A}$, we identify the state that is reachable via $s_A$. Let the outgoing transitions from this state be $s_{B_1},\ldots,s_{B_n}$. Let $S=\{(s_{A},s_{B_1}),\ldots,(s_{A},s_{B_n})\}$. A suitable common ABC is searched for all the transition pairs in a set $S_a \subseteq S$. For every pair $(s_A, s_{B_j}) \in S_a$, for all $0 \leq j \leq n$, we also have that $(s_A,s_{B_j}) \in \mathcal{S}(\bfm{r}_i)$ for some $0 \leq i \leq k$. If such a common ABC exists, then such lassos $\bfm{r}_i$ can be discarded from further consideration as existence of ABCs disallows the transitions in those lassos and are collected in the set $\mathcal{R}_d$. Note that, unfortunately, there is no systematic way to obtain the set $S_a\subseteq S$ consisting of all transition pairs that admit a common ABC. $S_a$ is first picked in a trial-and-error fashion and then an \emph{oracle} (c.f Remark~\ref{remark:Complexity}) is used to determine if the transition pairs in $S_a$ admit a common ABC.

This procedure is then repeated for every transition label at the initial state, and the discarded lassos are iteratively added to $\mathcal{R}_d$. Once all the outgoing transition labels are covered, we move on to the next state reachable from the initial state and repeat the procedure to find ABCs for only those pairs that belong to $\mathcal{S}(\bfm{r}_i)$ such that $\bfm{r}_i \in \mathcal{R} \backslash \mathcal{R}_d$. This continues in a breadth-first search fashion until all the lassos are discarded, \textit{i.e.}, $\mathcal{R}_d=\mathcal{R}$, in which case we can conclude that $T(\mathfrak{S},L) \models \phi$, or all the states of $\mathcal{A}_{\neg \psi}$ have been considered. If $\mathcal{R}_d \subset \mathcal{R}$, it means that there are lassos for which no ABC could be found, rendering the verification procedure inconclusive.  

\begin{remark}
It is important to note that this procedure is valid under the assumption that $\mathcal{A}_{\neg \psi}$ can be represented as a DBA. Unfortunately, there may be cases where this is not possible as DBA are strictly less expressive than NBA. However, Algorithms~\ref{alg:AEprocedure} and \ref{alg:ABC_FIND} may be adapted for any arbitrary specifications in which $\mathcal{A}_{\neg \psi}$ cannot be expressed as a DBA by considering $\omega$-automata with other acceptance conditions such as Rabin~\cite{baier2008principles} that are closed under nondeterministic choice. This construction is provided in the technical appendix. 
\end{remark}

\begin{lemma}[Beyond determinisitc B\"uchi]\label{lem:AE_arbitrary}
 Given a HyperLTL specification $\phi = \forall \pi_1 \ldots \forall \pi_l \exists \pi_{l+1} \ldots  \exists \pi_p \psi$ such that $\neg \psi$ is an NBA, Algorithms~\ref{alg:AEprocedure} and \ref{alg:ABC_FIND}  can be extended to work with deterministic Rabin automata characterizing $\neg \psi$. 
\end{lemma}

\begin{algorithm}[h!]
\begin{algorithmic}
\Require{$\mathfrak{S},\phi= \forall \pi_1 \ldots \forall \pi_{l} \exists \pi_{l+1} \ldots \exists \pi_p \psi, L $ }
\State Construct DBA $\mathcal{A}_{\neg \psi}$ for $\neg \psi$
\State Identify lassos $\mathcal{R} := \{\bfm{r}_1,\bfm{r}_2,\ldots,\bfm{r}_k\}$ of $\mathcal{A}_{\neg \psi}$ 
\State $R_d \gets \emptyset$
\State $\mathcal{A}'_{\neg \psi}:=(\Sigma^p,Q',q_0,\Delta',F') \gets Prune(\mathcal{A}_{\neg \psi})$ 
\State $visited \gets [0, \ldots 0]$ \Comment{Array of size $|Q'|$}
\State $V \gets \{q_0\}$
\State $visited[q_0] \gets 1$
\While{ $V \neq \emptyset $}
    \For{\textbf{each} $r \in V$}
        \State $\mathcal{R}_d \gets \mathcal{R}_d \cup ABC\_FIND(\mathcal{A}'_{\neg \psi},r)$ 
        \State $\mathcal{R}_m  \gets \mathcal{R} \setminus \mathcal{R}_d$
        \If{$\mathcal{R}_m = \emptyset$} 
        \State \textbf{return} $T(\mathfrak{S},L) \models \phi$ 
        \EndIf
        \For{\textbf{each} $s_A \in \Sigma^p$}
        \State $r' \gets \Delta'(r,s_A)$
        \State $ G \gets \{\bfm{r} \in \mathcal{R}_m \mid (r,r') \in \bfm{r} \}$
        \If{$G \neq \emptyset$ and $visited[r'] < 1$}
        \State $V \gets V \cup \{ r' \}$
        \State $visited[r'] \gets visited[r'] + 1$
        \EndIf
        \EndFor
        \State $V \gets V \setminus \{ r \}$
    \EndFor
\EndWhile
\State \textbf{return} Inconclusive

\end{algorithmic}
\caption{Algorithm for verification of $\forall^{*} \exists^{*}$ fragment of HyperLTL}
\label{alg:AEprocedure}
\end{algorithm}

\begin{algorithm}[h!]
\begin{algorithmic}
\Require{$\mathcal{A}'_{\neg \psi} := (\Sigma^p,Q',q_0,\Delta',F'), r$}
\State $R_d \gets \emptyset$
\For{\textbf{each} $s_A \in \Sigma^p$}
    \State $S \gets \emptyset$
    \State $r' \gets \Delta'(r,s_A)$
    \For{\textbf{each} $s_B \in \Sigma^p$}
        \If{$\Delta'(r',s_B) \neq \emptyset$}
        $S \gets S \cup \{ (s_A,s_B) \}$
        \EndIf
    \EndFor
    \State Find a common ABC for a set $S_a \subseteq S$.
    \For{\textbf{each} $(s_A,s_B) \in S_a$}
    \State $r' \gets \Delta'(r,s_A)$
    \State $r'' \gets \Delta'(r',s_B)$
        \State $\mathcal{R}_d  \gets \mathcal{R}_d \cup \{\bfm{r} \in \mathcal{R} \mid (r,r',r'') \in \bfm{r} \}$ 
    \EndFor
\EndFor
\State \textbf{return} $\mathcal{R}_d$
\end{algorithmic}
\caption{Function $ABC\_FIND$}
\label{alg:ABC_FIND}
\end{algorithm}

%% file: synthesis.tex
In the previous sections, we showed that existence of an ABC satisfying conditions \eqref{eq:bar1}-\eqref{eq:bar3} for a transition pair $(s_A,s_B) $ is vital to verify that a \dtCS $\mathfrak{S}$ satisfies a desired HyperLTL specification $\phi$. In general, synthesizing such ABCs is a difficult problem. However, under some assumptions on the type of ABCs considered, the dynamics of the systems, and the geometry of the state sets, one can efficiently compute ABCs that can sufficiently prove the satisfaction of conditional invariance guarantees. This can in turn be utilized to verify the satisfaction of HyperLTL specifications. Specifically, we see that when the dynamics of the systems are restricted to polynomial functions and the state set $X$, exogenous input set $W$ as well as the safe and unsafe sets obtained from $(s_A,s_B)$ are semi-algebraic sets \cite{bochnak_real_1998}, 
one can utilize sum-of-squares (SOS) programming techniques \cite{Parrilo2003} to compute polynomial ABCs of predefined degrees. We now formally state the following assumption. 

\begin{assumption} \label{as:sos}
The \dtCS $\mathfrak{S}$ has a continuous state set $X \subseteq \R^{n}$ and continuous exogenous input set $W \in \R^m$, and its transition function $f: X \times W \rightarrow X$ is a polynomial function of the state $x$ and input $w$.
\end{assumption}

Under Assumption \ref{as:sos}, one can readily observe that the state and input sets of augmented \dtCS $\mathfrak{S}^p$ (i.e. $X^p$ and $W^p$, respectively) are also continuous, and the function $f^p: X^p \times W^p \rightarrow X^p$ is a $p$-tuple of polynomial functions.  Having this, one can then reformulate conditions \eqref{eq:bar1}-\eqref{eq:bar3} as an SOS optimization problem (cf. next lemma) to search for a polynomial ABC for augmented \dtCS $\mathfrak{S}^p$. In order to present the result below, we assume that the number of quantifiers ``$\exists$" in $\phi=\mu_1 \pi_1\ldots \mu_p \pi_p \psi$ is equal to $k$ and define $I_{\exists} = \{i \mid \mu_i=\exists, 1\leq i \leq p\}$. 

\begin{lemma} \label{lem:sos}
Suppose Assumption \ref{as:sos} holds and sets $X^p$, $A$, $B$, and $W^p$ are defined as $X^p=\{\tilde x \in \mathbb{R}^{np} \mid g(\tilde x) \geq 0\}$, $A=\{\tilde x \in \mathbb{R}^{np} \mid g_{0}(\tilde x) \geq 0\}$, $B=\{\tilde x \in \mathbb{R}^{np} \mid g_{u}(\tilde x) \geq 0\}$, and $W^p = \{ \tilde w \in \mathbb{R}^{mp} \mid g_{in}(\tilde w) \geq 0 \}$, where the inequalities are considered component-wise and functions $g, g_0, g_u$, and $g_{in}$ are polynomials. 
Suppose there exist a polynomial $\mathbb B(\tilde x)$ and $k$ polynomials $h^{i}_{j}(\hat x_i, \hat w_i)$, $i \in I_{\exists}$, corresponding to the $j^{th}$ entry of $w_i=(w_{i_1},\ldots,w_{i_m})\in W \subseteq \mathbb{R}^m$, 
where $\hat x_i$ refers to those components of the state with indices less than $i$ and $\hat w_i$ denotes the inputs associated with ``$\forall$" quantifiers with indices less than $i$. In addition, suppose there exist sum-of-squares polynomials $\lambda(\tilde x, \tilde w), \lambda_0(\tilde x), \lambda_u(\tilde x)$, and $\lambda_{in}(\tilde x,\tilde w)$ of appropriate dimensions,  such that the following expressions are sum-of-square polynomials:
\begin{align}
  - & \B(\tilde x) -\lambda_0(\tilde x)g_0^T(\tilde x), \label{eq:sos1} \\
  & \B(\tilde x) - \lambda_u(\tilde x)g_u^T(\tilde x)-\varepsilon, \label{eq:sos2} \\
  - & \B(f^p(\tilde x, \tilde w)) + \B(\tilde x) - \lambda(\tilde x, \tilde w)g^T(\tilde x) -\lambda_{in}(\tilde x, \tilde w)g_{in}^T(\tilde w) -\sum_{i\in I_{\exists}} \sum_{j=1}^m (w_{i_j}-h^i_j(\hat x_i, \hat w_i)),  \label{eq:sos3}
\end{align}
where $\varepsilon$ is a small positive number. Then, $\B(\tilde x)$ is an ABC from set $A$ to set $B$ satisfying conditions \eqref{eq:bar1}-\eqref{eq:bar3}. 
Note that a small tolerance $\varepsilon$ is needed in \eqref{eq:sos2} to ensure strict positivity of ABC as required in \eqref{eq:bar2}. 
\end{lemma}

Existing tools such as SOSTools \cite{papachristodoulou2013sostools} can be used in conjunction with semi-definite programming solvers such as SeDuMi \cite{sturm1999using} to compute polynomial ABCs satisfying \eqref{eq:sos1}-\eqref{eq:sos3}. 

\begin{remark}
Note that the SOS approach for computing barrier certificates is restricted to systems with polynomial-type dynamics. One can also utilize different computational techniques when dynamics are not necessarily polynomial. For example, conditions \eqref{eq:bar1}-\eqref{eq:bar3} can be reformulated as a satisfiability problem and SMT solvers such as Z3 \cite{de2008z3} and dReal \cite{gao-complete_2012} can be utilized to search for suitable barrier certificates using counterexample-guided inductive synthesis framework \cite{jagtap_formal_2019}. One can also train suitable barrier certificates using neural networks e.g. \cite{peruffo2020automated,barrier_nn}. 
\end{remark}

%% file: casestudies_new.tex
\subsection{Approximate Initial-State Opacity of a Vehicle Model} 
\label{casestudy:1}

\begin{figure*}[t]
\vspace{-0.7em}
\centering
\begin{subfigure}{0.33\textwidth}
\centering
\begin{tikzpicture}[node distance=2.5cm,thick,
 el/.style = {inner sep=1pt, align=left},
 every label/.append style = {font=\tiny},
 every  edge/.append style = {draw, -stealth', shorten > = 1pt,
                              font=\footnotesize, inner sep=2pt,auto}]
    
     \node[initial, state, initial text=, ] (0) {$q_0$};
     \node[state, below right of = 0, ] (1) {$q_1$};
     \node[state, accepting, above right of = 1, ] (2) {$q_2$};
     \path [->]
     (0) edge node{$(a_1,a_1)$} (2)
     (0) edge  node[xshift = -2.75cm, pos = 0.75] {$(a_1,a_2) \wedge (a_3,a_3)$} (1) 
     (1) edge node[above,xshift = 1cm, pos = 0.1]{$(a_4,a_4)$} (2) 
     (1) edge[loop below] node{$(a_3,a_3)$} (1)
     (2) edge [loop above] node{$\top$} (2);
     \end{tikzpicture}
\caption{}
\label{subfig:spec}
\end{subfigure}
\begin{subfigure}{0.33\textwidth}
\includegraphics[width=1\textwidth]{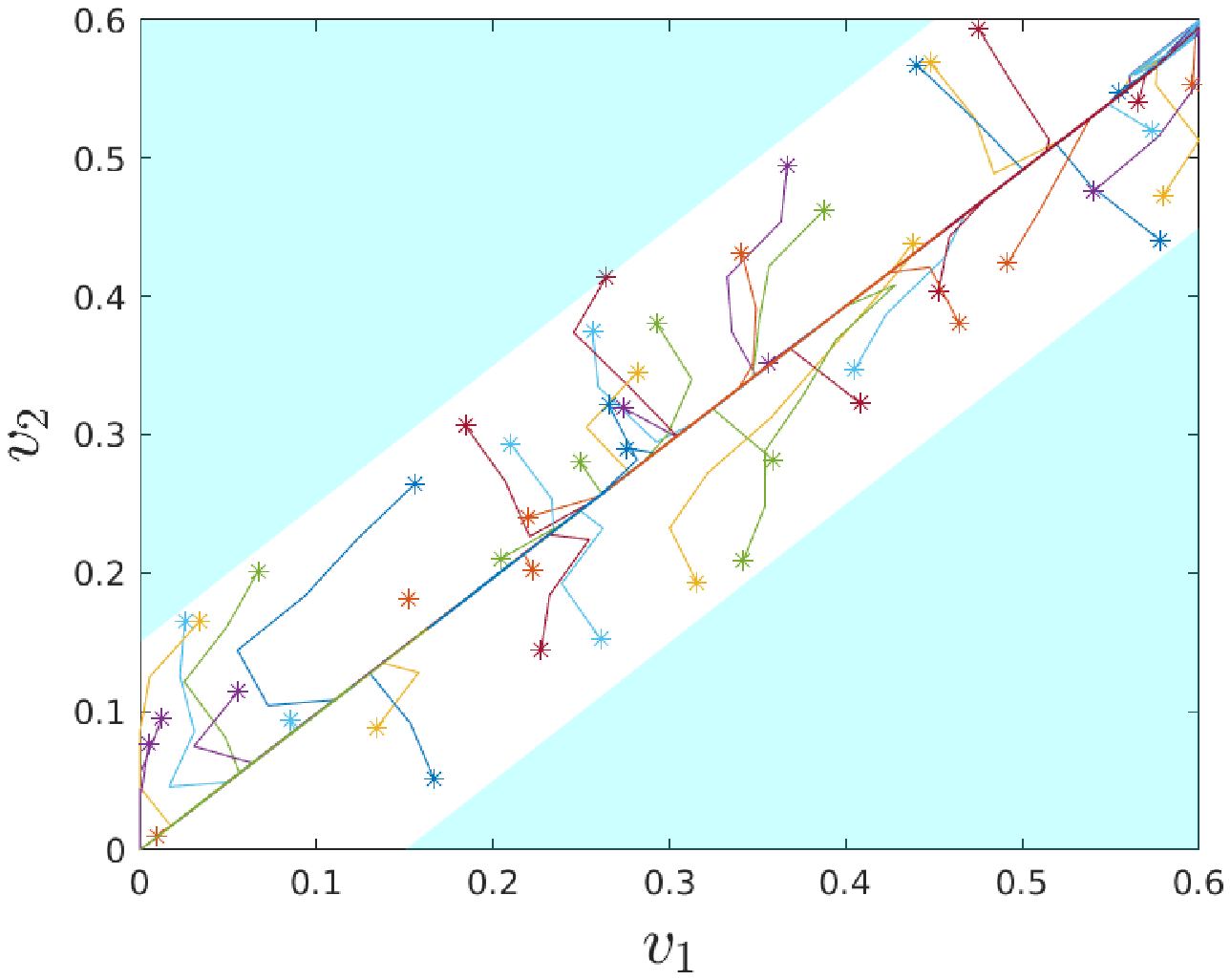}
\centering
\caption{}
\label{subfig:simu}
\end{subfigure}
\begin{subfigure}{0.33\textwidth}
\includegraphics[width=1\textwidth]{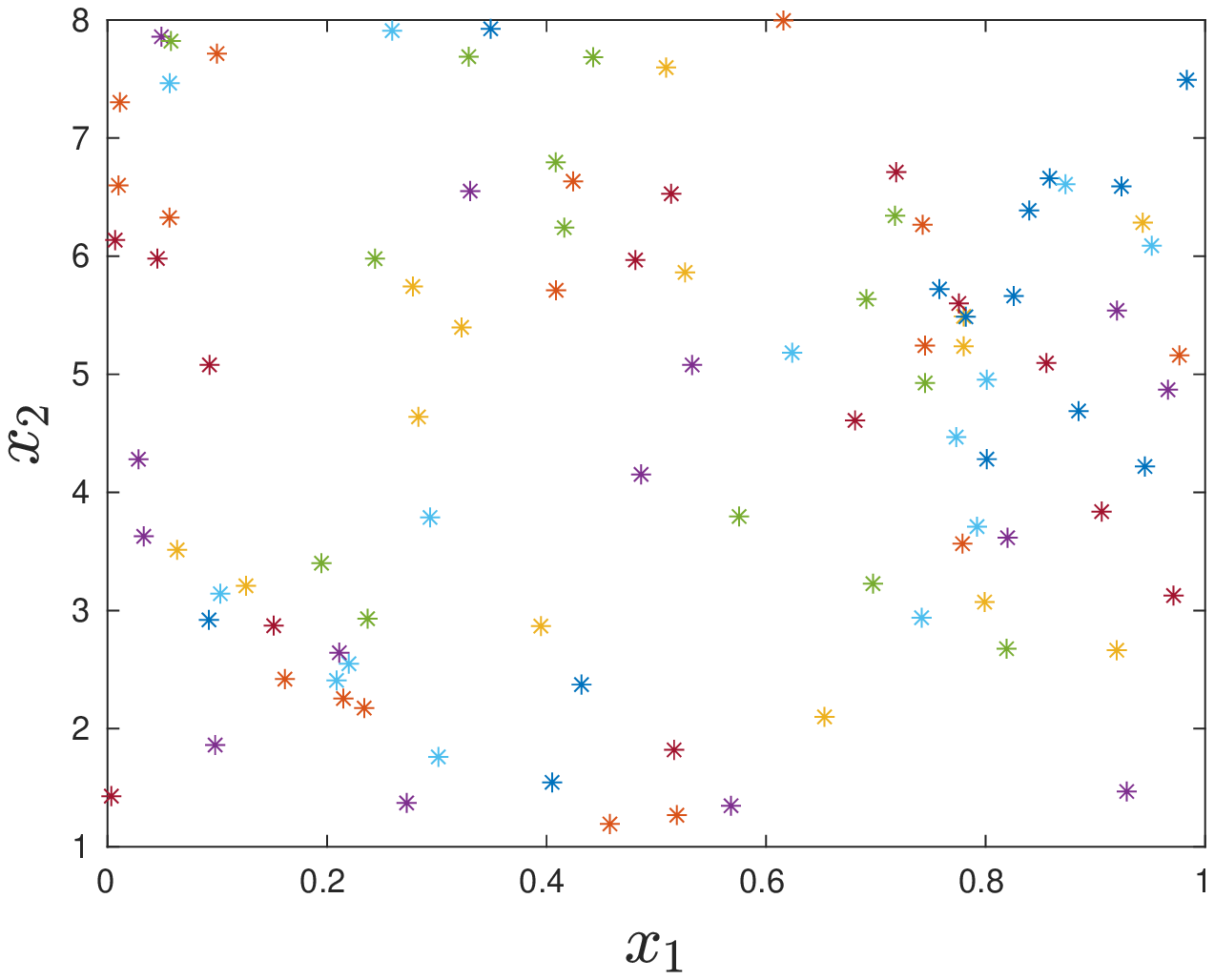} 
\centering
\caption{}
\label{subfig:simu_x}
\end{subfigure}
\caption{ (a) NBA $\mathcal{A}_{\neg \psi}$ corresponding to $\neg \psi$ (b) State runs of $\mathfrak{S}^2$ projected over the velocity coordinate. Region in blue indicates the unsafe set (c) The initial conditions of the state runs projected over the position coordinate marked by \textasteriskcentered{} which show that the first initial condition (i.e. $x_1$) is secret and the other one (i.e. $x_2$) is non-secret.}
\label{fig:simu}
\vspace{1em}
\end{figure*}

In this example, we consider 
the discrete-time, two-dimensional model of an autonomous vehicle on a single-lane road, with state variables as $x=[s,v]$, where $s$ denotes the absolute position of the vehicle and $v$ denotes the absolute velocity. The dynamics of the system are borrowed from \cite{liu} and governed by:
\begin{equation} \label{eq:veh-dyn}
 \mathfrak{S}: \begin{cases} & s(t+1)= s(t)+\Delta \tau v(t)+ \frac{\Delta \tau^2}{2} w(t), \\
 & v(t+1)=v(t)+\Delta \tau w(t),
 \end{cases}
\end{equation}
where $w$ is the exogenous input, \textit{i.e.}, acceleration, and $\Delta \tau=1$ is the sampling time. Here, we verify the $\delta$-approximate initial state opacity property \cite{liu} for this system.  The specification requires that, for any state run of the system that begins from a secret state, there must exist another state run that begins from a non-secret state such that both state runs render $\delta$-close observations from the observer's (or intruder's) point of view.  The significance of the specification can be motivated with the help of a simple example. Consider a scenario where the vehicle is assigned for a cash transit from a high-security bank to an ATM machine, and the initial locations of the vehicle must be kept secret. It is assumed that a malicious intruder is observing the velocity of the vehicle remotely and intends to gain access to the secret information and perform an attack. Therefore, it is critical to ensure that the secret states of the system are never revealed to the intruder. This security specification can be modeled as a $\delta$-approximate initial-state opacity problem, where $\delta \geq 0$ captures the measurement precision of the intruder. 

To express $\delta$-approximate initial-state opacity as a HyperLTL specification, consider system \eqref{eq:veh-dyn} with state set $X=[0,8] \times [0,0.6]$ and exogenous input set $W=[-0.04,0.04]$. The secret set is defined by $X_1=[0,1] \times [0,0.6]$ and the non-secret set is consequently given by $X_2=X \backslash X_1$. 
Here, we assume that the intruder can only observe the velocity of the car with a precision of $\delta$, \textit{i.e}, observations of two states $x_1=[s_1,v_1] \in X$ and $x_2=[s_2,v_2] \in X$ appear identical to the intruder if $\|v_1-v_2\| \leq \delta$. We now construct atomic propositions as $\AP=\{a_1,a_2,a_3,a_4\}$ where $a_1$ and $a_2$ are such that $L(x \in X_z)=a_z$ for $z=\{1,2\}$. The atomic propositions $a_3$ and $a_4$ are constructed over the augmented state set such that we have $(a_{3},a_{3}) := \{(L(x_1{=}[s_1,v_1] \in X),L(x_2{=}[s_2,v_2] \in X)) \mid \|v_1-v_2\|^2 \leq \delta^2\}$ and  $(a_{4},a_{4}) := \{(L(x_1{=}[s_1,v_1] \in X),L(x_2{=}s_2,v_2] \in X)) \mid \|v_1-v_2\|^2 \geq \delta^2+\epsilon\}$, where $\epsilon$ is a small positive number introduced to certify positivity using SOS programming. Note that atomic propositions for HyperLTL specifications are usually defined over the single system rather than the augmented one. On the other hand, the $\delta$-approximate initial state opacity specification requires the atomic propositions to capture the $\delta$-closeness between any two states of the augmented system. In a finite-state system, one could quantify  $\delta$-closeness by using finite conjuncts of atomic propositions defined over the original system, but in the infinite-state case such as ours, that is not possible. Therefore, to handle this non-trivial case, we modify atomic propositions slightly and define them over the augmented state set. Such modifications can be made without any loss of generality in our approach.
Now, one can formulate the $\delta$-approximate initial-state opacity specification as a HyperLTL formula given by $\phi = \forall \pi_1 \exists \pi_2 \psi$, where $\psi=a_{1\pi_1} \rightarrow (a_{2\pi_2} \wedge \always(a_{3\pi_1} \wedge a_{3\pi_2}))$.  

Consider the system $\mathfrak{S}^2{=}\mathfrak{S} \times \mathfrak{S}$ with states $(x_1{=}[s_1,v_1],x_2{=}[s_2,v_2]) \in X^2$ and input $(w_1,w_2) \in W^2$, and the NBA $\mathcal{A}_{\neg \psi}$ corresponding to $\neg \psi$ that is obtained as shown in Figure \ref{subfig:spec}. 
We decompose $\mathcal{A}_{\neg\psi}$ to obtain transition pairs for all lassos. This is obtained as $((a_1,a_2) \wedge (a_3,a_3),(a_4,a_4))$, $((a_1,a_1),\top)$ and $((a_4,a_4),\top)$. The latter two do not admit ABC following Remark \ref{remarkdecomp}, and the transition pair $((a_1,a_1),\top)$ is ignored by assuming that the augmented system $\mathfrak{S}^2$ never starts from an initial condition corresponding to $\tilde a_1=(a_1, a_1)$. Note that this assumption is only on the $\textit{virtual}$ copy of the system $\mathfrak{S}$ 
and does not restrict the initial states of the original system $\mathfrak{S}$ directly. For the transition pair $((a_1,a_2) \wedge (a_3,a_3),(a_4,a_4))$, we compute a suitable ABC by considering $\delta=0.15$. Using SOSTOOLS and SeDuMi tools on MATLAB, and with tolerance parameters $\epsilon=0.01$ and $\varepsilon=0.01$, we obtain ABC as follows.
\begin{align*}
 &\mathbb{B}((s_1,v_1),(s_2,v_2))= 58.76v_1^2 - 117.7v_1v_2 + 0.003644v_1s_1 \\ &- 0.004372v_1s_2 + 0.02869v_1 + 58.91v_2^2 - 0.003608v_2s_1 \nonumber \\
 & + 0.004257v_2s_2 - 0.008283v_2 +  0.004117s_1^2 - 0.008284s_1s_2 \nonumber \\ & + 0.01649s_1 +  0.004272s_2^2 - 0.05732s_2 - 1.433, \nonumber
\end{align*}
and the corresponding $\exists$ quantifier on the input is fulfilled by $w_2(s_1,v_1,s_2,v_2,w_1)$ $\!=\!  0.983v_1^2$ $-v_2^2+w_1$. 
Therefore, we conclude that the system $\mathfrak{S}$ satisfies the HyperLTL specification $\phi$ representing $\delta$-approximate initial-state opacity problem with $\delta=0.15$. Figure \ref{subfig:simu} shows the projection of a few state runs on the velocity coordinate of the augmented system $\mathfrak{S}$, with initial conditions in $A={L^p}^{-1}((a_1,a_2) \wedge (a_3,a_3))$. Figure \ref{subfig:simu_x} shows the initial conditions projected on the position coordinate. It follows that the state runs avoid reaching the unsafe regions, indicating that the original system is $\delta$-approximate initial-state opacity. 
We should add that the computation of ABCs using the mentioned tools on MATLAB takes roughly $35$ seconds on a machine running with Linux Ubuntu OS (Intel i7-8665U CPU with a $32$ GB of RAM).
\vspace{-1em}

\subsection{Initial-State Robustness of a Room Temperature Example} 

In this example, we verify the initial-state robustness of a safety controller obtained for a single room temperature system, borrowed from \cite{jagtap_formal_2019}. The discrete-time evolution of the temperature $T(\cdot)$ in the presence of a safety controller (as designed in \cite{jagtap_compositional_2020}) 
is given by
\begin{equation} \label{eq:temp-dyn}
\mathfrak{S}:  T(k+1)= T(k)+\tau_s\alpha_e(T_e-T(k)) +\tau_s\alpha_h(T_h-T(k))( c_1T(k) + c_2),
\end{equation}
where parameters $\alpha_e=0.008$ and $\alpha_h=0.0036$ are heat exchange coefficients, $T_e= 15^{\circ} C$ is the ambient temperature, $T_h=55^{\circ} C$ is the heater temperature, $c_1 = -0.0024$ and $c_2 = 0.5357$ are controller parameters, and $\tau_s= 5$ minutes is the sampling time. The initial-state robustness specification requires that if a state run started from a given initial condition remains safe, then all the state runs starting from $\delta$-close initial conditions must also remain safe. Such a specification is especially useful when there are uncertainties arising from not knowing the exact initial state. Note that robustness is a commonly studied property in the classical control theory \cite{zhou_essentials_1997}. However, we provide here an alternate method to verify robustness by formulating it as a HyperLTL formula. 

To describe our specification as a HyperLTL formula, we consider the state set $X=[20,35]$. We further introduce the safe set as $X_1=[20,25]$ and the unsafe set as $X_2=[25,35]$. For the system $\mathfrak{S}$, the predefined initial state is given by $X_3=\{21\}$. We also define the set $X_4=[20.5,21.5]$ to capture $\delta$-close states with respect to the initial state, where $\delta=0.5$. The set of atomic propositions is $\AP=\{a_1,a_2,a_3,a_4\}$, where $L(x \in X_i)=a_i,$ for all $i \in \{1,2,3,4\}$. The HyperLTL formula for initial-state robustness specification is $\phi=\forall \pi_1 \forall \pi_2 (a_{3\pi_1} \wedge a_{4\pi_2}) \rightarrow \always (a_{1\pi_1} \wedge a_{1\pi_2})$. 

\begin{figure}[t] 
\centering
\begin{subfigure}{0.4\columnwidth}
\centering
\begin{tikzpicture}[node distance=1.75cm,thick,
 el/.style = {inner sep=1pt, align=left},
 every label/.append style = {font=\tiny},
 every  edge/.append style = {draw, -stealth', shorten > = 1pt,
                              font=\footnotesize, inner sep=2pt,auto}]
    
     \node[initial, state, initial text=,] (0) {$q_0$};
     \node[state, below of = 0, ] (1) {$q_1$};
     \node[state, accepting, below of = 1, ] (2) {$q_2$};
     \path [->]
     (0) edge  node {$(a_3,a_4)$} (1) 
     (1) edge node{$\neg (a_1,a_1)$} (2) 
     (2) edge [loop right] node{$\top$} (2);
     \end{tikzpicture}\caption{}
\label{subfig:spec2}
\end{subfigure}
\hspace{-0.5em}
\begin{subfigure}{0.5\columnwidth}
\centering
\includegraphics[width = 1\columnwidth]{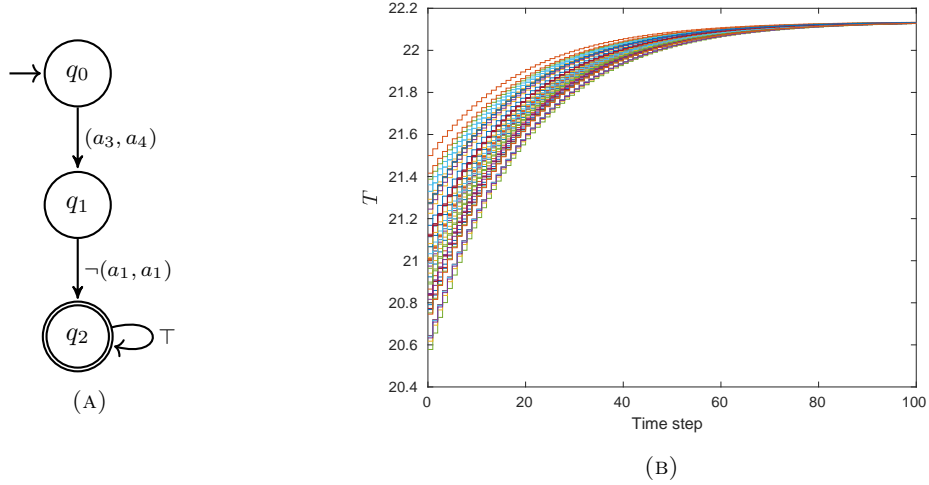}
\caption{}
\label{subfig:simu_rte}
\end{subfigure}
\vspace{-0.5em}
\caption{ (a) NBA $\mathcal{A}_{\neg \psi}$ corresponding to $\neg \psi$ and (b) state runs of the system $\mathfrak{S}$ starting from initial set $X_3$.} 
\end{figure}

Consider the augmented system $\mathfrak{S}^2=\mathfrak{S} \times \mathfrak{S}$ with states $(T_1,T_2)$ and the NBA $\mathcal{A}_{\neg \psi}$ corresponding to $\neg \psi$, which is obtained as shown in Figure \ref{subfig:spec2}. There is only one transition pair $((a_2,a_3),\neg(a_1,a_1))$. For this pair, we utilize SOS programming with the help of SOSTOOLS and SeDuMi to compute a polynomial ABC of degree $2$ as
$\B(T_1,T_2)=1.3402T_1^2 - 1.8137T_1T_2 - 19.6699T_1 + 1.2559T_2^2 - 15.8439T_2 + 399.7534$
with a tolerance of $\varepsilon=0.001$. The existence of the ABC proves that the safety controller designed for the system $\mathfrak{S}$, is indeed robust with respect to initial-state uncertainty with a robustness measure of $\delta=0.5$. Figure \ref{subfig:simu_rte} shows that the state runs obtained for the system $\mathfrak{S}$ remains in the safe set $X_1=[20,25]$ when starting from the initial set $X_3=[20.5,21.5]$ which captures uncertainties in the initial state. The time taken for the computation of ABC was around $19$ seconds on the same machine as in the first case study.

%% file: discussion.tex
\subsection{HyperLTL Synthesis}

An interesting problem that follows HyperLTL verification for dynamical systems is the synthesis of controllers ensuring the satisfaction of HyperLTL specifications. In this case, for a system $\mathfrak{S}=(X,U,f)$ and a HyperLTL specification $\phi$, one would view $\nu$ in~\eqref{eq:dyn} as control signal rather than exogenous one and design $\nu$ such that the corresponding traces $T$ of $\mathfrak{S}$ satisfy $\phi$. Unfortunately,
there are major challenges in synthesizing controllers even when the HyperLTL specification is a simple conditional invariance (CI).
Let us consider a CI $\chi$, and a controller  $\mathcal{G}: X \rightarrow U$ such that $\nu(t) := \mathcal{G}(x(t))$. Then, condition \eqref{eq:bar3} of the barrier certificate for CI $\chi$ in the context of synthesis can be reformulated as: for any $\tilde x=(x_1, x_2, \ldots, x_p) \in X^p,$
\begin{align}
  \B(f(x_1, \mathcal{G}(x_1)), f(x_2, \mathcal{G}(x_2)), \ldots, f(x_p, \mathcal{G}(x_p))  )-\B(\tilde{x}) \leq 0. \label{eq:bar3_synth}
\end{align}
The above formulation ensures that the selection of the control input at any given state $x \in X$ according to map $\mathcal{G}$ is independent of the previous traces selected by the players.

However, to satisfy this condition, one must simultaneously search for suitable functions $\B$ and $\mathcal{G}$. This makes the above inequality non-convex in these unknown functions and unfortunately, one cannot leverage convex programming and correspondingly SOS and semi-definite programming to determine these functions even when they are assumed to be polynomials. However, given a map $\mathcal{G}$, one could search for a function $\B$ such that condition~\eqref{eq:bar3_synth} is satisfied, which is technically a verification problem and not a synthesis one anymore.
In general, even though one can verify whether a HyperLTL specification $\phi$ is realizable over a system, it is not possible to synthesize the control map $\mathcal{G}$ that ensures the satisfaction of $\phi$. In other words, it is not possible to find a solution to the HyperLTL synthesis problem. 

This is due to the fact that the inputs obtained satisfying condition~\eqref{eq:bar3} in the case of verification may depend on the previously quantified traces, which is not possible when considering controller synthesis. Remark that the problem of HyperLTL verification coincides with HyperLTL synthesis when the specification is of the form $\phi = \exists \pi_1\forall \pi_2 \ldots \forall \pi_p \psi $. To verify such specifications, it is sufficient to synthesize a controller for the first trace. Since the remaining traces are controlled by Abelard, ABCs satisfying condition~\eqref{eq:bar3} implies the satisfaction of condition~\eqref{eq:bar3_synth} with the controller $\mathcal{G}$ being the one synthesized for the first trace. We leave any further investigations for the synthesis problem as future work.

\subsection{Conclusion}

We proposed a discretization-free scheme for the verification of discrete-time uncertain dynamical systems against hyperproperties. 
Focusing on hyperproperties expressible as HyperLTL formulae, we presented an implicit automata-theoretic approach.
In our approach the specifications are reduced to a collection of conditional invariance properties by utilizing an implicitly quantified B\"uchi automata corresponding to the complements of the specifications. 
Working with an augmented system, we were able to devise a notion of augmented barrier certificates over the self-composition of the original system as a certificate of conditional invariance. 
The existence of ABCs is a sufficient proof that the conditional invariance holds, this provides a verification guarantee over the satisfaction of the hyperproperty. For a general HyperLTL specification, we showed that a common ABC for at least one conditional invariance in every lasso is required to provide verification guarantees. However, for a HyperLTL specification in the $\forall^{*}\exists^{*}$ fragment, we provided a systematic algorithmic procedure that leverages the structure of the automata to allow for different ABCs for different lassos.
We exploited a sum-of-squares approach to efficiently compute suitable ABCs. As future work, we plan on investigating approaches that allow for multiple ABCs to guarantee the satisfaction of general HyperLTL specifications. We would also investigate verification problems for stochastic systems and
synthesis problems against hyperproperties for continuous-state control systems. Moreover, we would also utilize compositionality approaches \cite{jagtap_compositional_2020} to tackle scalability issues in computing ABCs for large-scale systems.

\vspace{-1em}

%% file: appendix.tex
\subsection{Proof of Lemma \ref{lem:weaken}}
\label{appendix:prooflemweak}
If Eloise has a positional winning strategy in the turn-based evaluation game, then she can use the 
same strategy to choose traces in the stage-based evaluation game such that each index depends only on the states and 
actions at the current index in the traces quantified so far. Then, against an arbitrary policy 
chosen by Abelard, the resulting $p$-tuple of of the traces satisfy the LTL specification $\psi$. 
\vspace{-0.5cm}
\subsection{Proof of Lemma~\ref{lem:ABCLogic}}
\label{appendix:LemmaABC_Logic_Proof}
We prove this by contradiction. 
Suppose an ABC exists for $\chi$, but $\chi$ is not valid. Then, regardless of Eloise's strategy, Abelard always has a strategy that allows him to win. Let the set of traces selected by the players be $T = \{\bm{\sigma}_1, \ldots \bm{\sigma}_p \}$. For Abelard to win, he must ensure that for some $j \in \mathbb{N} $, $\Pi[j, \infty] \models_T s_A$ and for some $k > j$, $\Pi[k, \infty] \models_T s_B$ (from the existence of an ABC and Remark~\ref{remarkdecomp} we can conclude that $k \neq j$) to falsify $\always(s_A \rightarrow \always(\neg s_B))$. We consider the case where the strategy of Eloise is to select inputs according to condition \eqref{eq:bar3}. We note that selecting such a strategy leads to a non-increase in the value of the barrier certificate for the corresponding state in the augmented system, regardless of Abelard's strategy.
Let the set of traces at positions $j$ and $k$ correspond to states $\tilde x$ and $\tilde x'$ in the augmented system and let the corresponding input sequence that takes us from $\tilde x$ to $\tilde x'$ be $\tilde{\nu}$. 
From conditions \eqref{eq:bar1} and \eqref{eq:bar2}, we have $\B(\tilde x) \leq 0$ and $\B(\tilde{x}') > 0$. 
For any $l \geq 0$, let $\tilde w = \tilde{\nu}(l)$, and $\tilde x_l = \bfm{x}(l)$,  then we have $\B(f^p(\tilde x_l,\tilde w)) \leq \B(\tilde x_l)$ from condition \eqref{eq:bar3} regardless of Abelard's strategy. By induction on this condition, we can infer that $\B(\tilde{x}') \leq 0$. This is a contradiction to condition  \eqref{eq:bar2}. So, we infer that $\chi$ is valid in the turn-based game setting. Therefore, $\chi$ is also valid in the stage-based game setting according to Lemma \ref{lem:weaken}.

\vspace{-0.5cm}
\subsection{Proof of Lemma~\ref{lem:DisjunctABC}}
From Lemma \ref{lem:ABCLogic}, for a conditional invariance $\chi_j=\mu_1 \pi_1 \ldots \mu_p \pi_p \always(s_{A_j} \rightarrow \always( \neg s_{B_j}))$, existence of an ABC $\B_j$ implies that Eloise has a winning strategy to ensure that $T(\mathfrak{S},L) \models \chi_j$. Therefore, for a set of conditional invariances $\chi_1, \ldots, \chi_k$, Eloise may choose the same winning strategy corresponding to $\chi_j$ to ensure that at least one of the conditional invariances in the set holds. Therefore, we get $T(\mathfrak{S},L) \models \chi$, where $\chi= \mu_1 \pi_1 \ldots \mu_p \pi_p \underset{1 \leq j \leq k}{\bigvee} \always(s_{A_j} \rightarrow \always( \neg s_{B_j})).$

\vspace{-0.5cm}

\subsection{Proof of Lemma~\ref{lem:Common_ABC}}
The proof once again follows from Lemma~\ref{lem:ABCLogic}. The existence of a common ABC $\B$ guarantees that condition \eqref{eq:bar3} is satisfied for all conditional invariances $\chi_i$, $1 \leq i \leq k$. This implies that Eloise may use the same strategy to disallow all the transition pairs $(s_{A_i},s_{B_i})$, $1 \leq i \leq k$.  Therefore, we have $T(\mathfrak{S},L) \models \chi$, where $\chi= \mu_1 \pi_1 \ldots \mu_p \pi_p \underset{1 \leq i \leq k}{\bigwedge} \always(s_{A_i} \rightarrow \always( \neg s_{B_i}))$
\vspace{-0.5cm}
\subsection{Proof of Lemma~\ref{lem:ConjunctDisjunctABC}}
From Lemma \ref{lem:DisjunctABC}, for the set of conditional invariances $\{\chi_{i,1},\ldots,\chi_{i,v_i}\}$ for some $1 \leq i \leq k$, the existence of ABC $\B$ for some $\chi_{i,j}$ implies that $T(\mathfrak{S},L) \models \mu_1 \pi_1 \ldots \mu_p \pi_p \underset{1 \leq j \leq v_i}{\bigvee} \always(s_{A_{i,j}} \rightarrow \always( \neg s_{B_{i,j}}))$. For each $1 \leq i \leq k$, if there exists a common ABC $\B$ for some $\chi_{i,j}$, $1 \leq j \leq v_i$, then by Lemma \ref{lem:Common_ABC}, we have that $T(\mathfrak{S},L) \models \mu_1 \pi_1 \ldots \mu_p \pi_p \underset{1 \leq i \leq k}{\bigwedge} \always(s_{A_{i,j}} \rightarrow \always( \neg s_{B_{i,j}}))$. By combining these two results, for a family of set of conditional invariances $\{\{\chi_{1,1},\ldots,\chi_{1,v_1}\},\ldots,\{\chi_{k,1},\ldots,\chi_{k,v_k}\}\}$, one has $T(\mathfrak{S},L) \models \chi$, where $\chi = \mu_1 \pi_1 \ldots \mu_p \pi_p  \underset{(1 \leq i \leq k) }{\bigwedge } \underset{ (1 \leq j \leq v_{i}) }{\bigvee }\always (s_{A_{i,j}} \rightarrow \always(\neg s_{B_{i,j}}))$. 

\vspace{-0.5cm}
\subsection{Proof of Theorem~\ref{thm:main_result}}
Let the NBA $\mathcal{A}_{\neg \psi}$ corresponding to $\neg \psi$ have $k$ lassos, such that  the $i^{th}$ lasso has $v_i$ pairs of consecutive transitions. Let the pair $(s_{A_{i,j}}, s_{B_{i,j}})$ correspond to the $j^{th}$ pair of consecutive transitions along the $i^{th}$ lasso. Let $\chi_{i,j} = \mu_1 \pi_1 \ldots \mu_p \pi_p \always(s_{A_{i,j}} \rightarrow \always( \neg s_{B_{i,j}}))$ denote a conditional invariance specification and let us consider the set of conditonal invariances $\{ \chi_{1,1}, \ldots, \chi_{1,v_{1}}, \chi_{2,1} \ldots \chi_{2,v_{2}},\ldots, \chi_{k,1}, \ldots \chi_{k,v_{k}} \}$.  Then the existence of a common ABC satisfying Lemma~\ref{lem:ConjunctDisjunctABC} implies that if the augmented system lands on a state satisfying $s_{A_{i,j}}$, Eloise has a strategy to ensure that it never reaches a state satisfying $s_{B_{i,j}}$ for every $1 \leq i \leq k$, and some $1 \leq j \leq v_j$. However, to satisfy $\neg \psi$, Abelard must have a strategy that allows him to visit a state satisfying some $s_{A_{i,j}}$ and then later visit a state satisfying $s_{B_{i,j}}$, for some $1 \leq i \leq k$ and every $1 \leq j \leq v_i$, to follow the transitions along the $i^{th}$ lasso. Since this is not possible due to the existence of the ABC, we infer that $\phi$ is satisfied.

\subsection{Proof of Algorithms~\ref{alg:AEprocedure} and \ref{alg:ABC_FIND}}
Algorithm~\ref{alg:AEprocedure} is a sound way of obtaining suitable common ABCs in the presence of two or more outgoing edges from any state in DBA $\mathcal{A}_{\neg \psi}$, and different ABCs otherwise. It calls on function $ABC\_FIND$, presented in Algorithm \ref{alg:ABC_FIND}, that takes a state $r$ in $\mathcal{A}_{\neg \psi}$ as input and returns a set of lassos that pass through $r$ and can be denied by means of a common ABC. The correctness of Algorithm~\ref{alg:ABC_FIND} can be established by ensuring Eloise has a strategy to disallow all the lassos in $\mathcal{R}_d$. Initially, we have $\mathcal{R}_d = \emptyset$ and therefore Eloise may use any strategy. Now, at some iteration of the outermost loop, let $\mathcal{R}_d$ consist of the lassos that have already been disallowed. Furthermore, consider an outgoing label $s_A \in \Sigma^p$ from $r$ to $r'$. The first inner loop then identifies the labels $s_B$ such that $\Delta'(r',s_B) \neq \emptyset$, and constructs a set $S$ that consists of all such pairs $(s_A,s_B)$. The algorithm then finds a common ABC for the set $S_a \subseteq S$ and, by Lemma~\ref{lem:Common_ABC}, we can infer that Eloise has a strategy of selecting inputs such that she can always avoid those states that satisfy $s_{B}$ in the augmented system after satisfying $s_A$. Therefore, at any iteration of the second inner loop, Eloise has a strategy to avoid the lassos already present in $\mathcal{R}_d$, or the lassos corresponding to the pairs $(s_A,s_B) \in S_a$. 
Moreover, due to the specification being in the $\forall^{*}\exists^{*}$ fragment and the determinism in the automata, Eloise can uniquely determine the state of the augmented system as well as the current state in the automaton. This allows Eloise to select a unique strategy to avoid lassos $\bfm{r}$ that pass through the states $(r,r',r'')$  such that $r' = \Delta'(r,s_A)$ and $r'' = \Delta'(r',s_B)$, and add them to $\mathcal{R}_d$. Lastly, we note that if we can successfully find common ABCs then the algorithm terminates as there are finitely many labels in $\Sigma^p$.

We now prove the correctness of Algorithm~\ref{alg:AEprocedure} by using the fact that $\mathcal{R}_d$ consists of all the lassos that can be denied by Eloise. Initially, $\mathcal{R}_d$ is empty, and therefore, Eloise may use any strategy. Now, at some iteration of the outermost loop, consider a state $r$ and the set $\mathcal{R}_d$ that consists of the lassos that have already been denied. On calling the function $ABC\_FIND$ for a state $r$, we find a common ABC for those lassos $\bfm{r} \in \mathcal{R}$ such that $(r,r',r'') \in \bfm{r}$, where $\Delta'(r,s_A)=r'$, $\Delta'(r',s_B)=r'$ and $(s_A,s_B) \in S_a$. Then, from the proof of Algorithm~\ref{alg:ABC_FIND}, we have that Eloise has a unique strategy to deny those lassos.
If $\mathcal{R}_m = \emptyset$, then $\mathcal{R}_d = \mathcal{R}$. Thus all lassos that pass through some state in $F$ have been denied and the states in $F$ cannot be visited infinitely often. Then, we can infer that Eloise has a strategy to never satisfy $\neg \psi$ or that $T(\mathfrak{S},L) \models \phi$. However, if we traverse all the states and $R_m \neq \emptyset$, the algorithm is inconclusive.

\subsection{Proof of Lemma~\ref{lem:AE_arbitrary}}
An equivalent representation of an arbitrary $\omega$-regular language is as a Deterministic Rabin Automaton (DRA). A deterministic Rabin automaton (DRA) is a tuple $\mathcal{A}_d=(\Sigma,Q,q_0,\Delta,F)$, where $\Sigma$ is the alphabet, $Q$ is a finite set of states, $\Delta: Q \times \Sigma \rightarrow Q$ is a deterministic transition function and $F=\{(G_1,B_1),\ldots,(G_h,B_h)\}$ is a set of pairs of sets of states.  A word $\textsc{w}$ is said to be accepted by the DRA if for the corresponding run $r$, we have that $\infi(r) \cap G_i \neq \emptyset$ and $\infi(r) \cap B_i = \emptyset$ for some $1 \leq i \leq h$. Thus given a HyperLTL specification $\phi = \forall \pi_1 \ldots \forall \pi_l \exists \pi_{l+1} \ldots \exists \pi_p \psi$, we construct a DRA $A_{d,\neg \psi}$ corresponding to $\neg \psi$  by utilizing standard LTL to DRA construction techniques~\cite{LTL_to_Rabin}. Similar to a DBA, in the case of DRA, Eloise is aware of the history of states visited as the automaton is deterministic. We then determine whether Eloise has a strategy to ensure that the acceptance condition of $\mathcal{A}_{d, \neg \psi}$ is violated. This can be done by disallowing either the traces of the system from visiting any state in the set $G_j$, for all $1 \leq j \leq h$, infinitely often, or by ensuring that all the states in $B_j$, for all $1 \leq j \leq h$, are visited infinitely often. To construct this condition in terms of conditional invariances, we consider the states in $G_j$ and obtain lassos that reach and cycle on some state in the sets $G_j$. Having this, a similar approach to the one provided in Section \ref{subsec:forallexists} can be used to compute a set of common and different ABCs to disallow these lassos, and conclude whether $T(\mathfrak{S},L) \models \phi$.

\subsection{Proof of Lemma~\ref{lem:sos}}
\label{proof:sos}
Since $\lambda_0(\tilde x)$ is an SOS polynomial, we have that $\lambda_0(\tilde x)g_0^T(\tilde x)$ is non-negative over $A$. Therefore, if condition \eqref{eq:sos1} is an SOS polynomial, and therefore non-negative, it would directly imply condition \eqref{eq:bar1}. Similarly, the SOS constraint \eqref{eq:sos2} implies condition \eqref{eq:bar2}. Now we show that condition \eqref{eq:sos3} implies \eqref{eq:bar3}. By selecting inputs $w_{i_j}=h^i_j(\hat x_i, \hat w)$, the last term in \eqref{eq:sos3} vanishes. Since the expression $\lambda(\tilde x, \tilde w)g^T(\tilde x)$ is non-negative over $X^p$ and $\lambda_{in}(\tilde x, \tilde w)g_{in}^T(\tilde w)$ is non-negative over $W^{p}$, we have that for all $\tilde x \in X^p$, $-\B(f^p(\tilde x, \tilde w)) + \B(\tilde x) \geq 0$. This implies that condition \eqref{eq:bar3} holds, thus concluding the proof.